\documentclass[prd,showpacs,amsmath,amssymb]{revtex4}

\usepackage{graphicx}
\usepackage{dcolumn}
\usepackage{bm}
\usepackage{psfrag}
\usepackage{subfigure}

\newcommand{\be}{\begin{eqnarray}}
\newcommand{\ee}{\end{eqnarray}}
\newcommand{\bn}{\begin{eqnarray*}}
\newcommand{\en}{\end{eqnarray*}}

\newcommand{\nn}{\nonumber \\}
\newcommand{\nl}{\\}

\renewcommand{\vec}[1]{\mbox{\boldmath$#1$}}

\renewcommand{\th}{\ensuremath{\theta}}
\newcommand{\vth}{\ensuremath{\vartheta}}
\newcommand{\Th}{\ensuremath{\Theta}}
\newcommand{\ph}{\ensuremath{\phi}}
\newcommand{\vph}{\ensuremath{\varphi}}
\newcommand{\al}{\ensuremath{\alpha}}
\newcommand{\bt}{\ensuremath{\beta}}
\newcommand{\sg}{\ensuremath{\sigma}}
\newcommand{\gm}{\ensuremath{\gamma}}
\newcommand{\dl}{\ensuremath{\delta}}
\newcommand{\lm}{\ensuremath{\lambda}}

\newcommand{\Dl}{\ensuremath{\Delta}}
\newcommand{\Sg}{\ensuremath{\Sigma}}
\newcommand{\Gm}{\ensuremath{\Gamma}}
\newcommand{\Om}{\ensuremath{\Omega}}
\newcommand{\OmK}{\ensuremath{\Omega_{\rm K}}}

\newcommand{\Cal}{\ensuremath{\hat{\al}}}
\newcommand{\Cbt}{\ensuremath{\hat{\bt}}}
\newcommand{\Cgm}{\ensuremath{\hat{\gm}}}
\newcommand{\Cdl}{\ensuremath{\hat{\dl}}}
\newcommand{\Ci}{\ensuremath{\hat{\imath}}}
\newcommand{\Cj}{\ensuremath{\hat{\jmath}}}
\newcommand{\Ck}{\ensuremath{\hat{k}}}
\newcommand{\Cl}{\ensuremath{\hat{l}}}

\newcommand{\ze}{\ensuremath{\hat{0}}}
\newcommand{\on}{\ensuremath{\hat{1}}}
\newcommand{\tw}{\ensuremath{\hat{2}}}
\newcommand{\tr}{\ensuremath{\hat{3}}}

\newcommand{\pvec}{\ensuremath{\vec{p}}}

\newcommand{\rvec}{\ensuremath{\vec{r}}}
\newcommand{\svec}{\ensuremath{\vec{s}}}

\newcommand{\Jvec}{\ensuremath{\vec{J}}}
\newcommand{\Vvec}{\ensuremath{\vec{V}}}
\newcommand{\Xvec}{\ensuremath{\vec{X}}}

\newcommand{\lt}{\ensuremath{\left}}
\newcommand{\rt}{\ensuremath{\right}}

\begin{document}

\pagenumbering{arabic}

\title{Dynamics of Extended Spinning Masses in a Gravitational Field}%

\author{Bahram Mashhoon}
\email{mashhoonb@missouri.edu}
\affiliation{%
Department of Physics and Astronomy, University of Missouri-Columbia \\
Columbia, Missouri, 65211, USA
}%
\author{Dinesh Singh}
\email{singhd@uregina.ca}
\affiliation{%
Department of Physics, University of Regina \\
Regina, Saskatchewan, S4S 0A2, Canada
}%
\date{\today}

\begin{abstract}

 We develop a first-order approximation method for the influence of spin on the motion of extended
spinning test masses in a gravitational field.
This approach is illustrated for approximately circular equatorial motion
in the exterior Kerr spacetime.
In this case, the analytic results for the first-order approximation are
compared to the numerical integration of the exact system and the
limitations of the first-order results are pointed out.
Furthermore, we employ our analytic results to illustrate the gravitomagnetic
clock effect for spinning particles.

\end{abstract}

\pacs{04.20.Cv}

\maketitle

\section{Introduction}
\label{Sec1}

Imagine the motion of an extended spinning body in the exterior vacuum region of an astronomical source.
Let $T^{\mu \nu}$ be the energy-momentum of the extended test mass; then, the motion is governed by the four equations
that are given by the dynamical law $T^{\mu \nu}{}_{; \, \nu} = 0$.
Consider a representative point inside the extended test body (the ``center of mass'') such that
$u_s^\mu = dx^\mu/d\sg$ is its four-velocity vector and $\sg$ is the proper time along its worldline.
The equations of motion can then be expressed relative to the chosen worldline.
This has been accomplished in an elegant manner by Dixon \cite{Dixon1}.
Using Synge's world function \cite{Synge}, Dixon has defined the infinite set of multipole moments of $T_{\mu \nu}$ in a way
that is qualitatively similar to the standard nonrelativistic theory and thereby expressed the equations of motion
as \cite{Dixon1}
%
\be
{DP^\mu \over d\sg} & = & - {1 \over 2} \, R^\mu{}_{\nu \alpha \beta} \, u_s^\nu \, S^{\alpha \beta} + {\cal F}^\mu,
\label{Dixon-p}
\nl
{DS^{\mu \nu} \over d\sg} & = & P^\mu \, u_s^\nu - P^\nu \, u_s^\mu + {\cal T}^{\mu \nu}.
\label{Dixon-s}
\ee
%
Here, the momentum vector $P^\mu$ and the spin tensor $S^{\mu \nu}$ are the first two moments of $T^{\mu \nu}$,
while the Dixon force ${\cal F}^\mu$ and torque ${\cal T}^{\mu \nu}$ are given in terms of the quadrupole and higher
moments of the test body.
It is necessary to add a supplementary condition to equations (\ref{Dixon-p})--(\ref{Dixon-s}) in order to fix an appropriate
center-of-mass trajectory.
It turns out that for an extended body the appropriate condition is \cite{Dixon2}
\be
S^{\mu \nu} \, P_\nu & = & 0.
\label{spin-condition}
\ee
The work of Dixon provides a natural generalization of the previous results of Mathisson \cite{Mathisson} and
Papapetrou \cite{Papapetrou} that were limited to pole-dipole particles.

In the present paper, we are interested in the dynamics of an extended body of mass $m$ and spin $s$ in the field
of a rotating mass $M \gg m$; therefore, we consider equations (\ref{Dixon-p})--(\ref{spin-condition}) without the quadrupole and
higher moments, i.e., ${\cal F}^\mu = 0$ and ${\cal T}^{\mu \nu} = 0$.
The resulting reduced Mathisson-Papapetrou-Dixon (``MPD'') equations then lead to natural definitions for the mass $m$
and spin $s$ of the extended test particle that are preserved throughout the motion.
We assume that the metric tensor has signature $+2$ and we use units such that $c = G = 1$, unless otherwise specified.
It then follows from Eqs.~(\ref{Dixon-p})--(\ref{spin-condition}) that in this case the mass of the extended particle is
given by $m = \lt(-P^\mu \, P_\mu\rt)^{1/2}$, which is a constant of the motion.
The spin vector is then defined by
\be
S_\mu & = & -{1 \over 2 \, m} \, \epsilon_{\mu \nu \rho \sg} \, P^\nu \, S^{\rho \sg}, \qquad
S^{\mu \nu} \ = \ {1 \over m} \, \epsilon^{\mu \nu \rho \sg} \, P_\rho \, S_\sg,
\label{S-vec}
\ee
where $\epsilon_{\mu \nu \rho \sg}$ is the Levi-Civita tensor given by $\epsilon_{\mu \nu \rho \sg} = \sqrt{-g} \, \sg_{\mu \nu \rho \sg}$
in terms of the alternating symbol with $\sg_{0123} = 1$.
It follows from the reduced MPD equations that $s^2 = S_\mu \, S^\mu = {1 \over 2} \, S_{\mu \nu} \, S^{\mu \nu}$ is a
constant of the motion.
Moreover, the requirement that the dipole force be much smaller than the monopole
force implies that $s \ll m \, r$, where $r$ is the distance between $m$ and $M$;
that is, the M{\o}ller radius of the extended test particle $\lt(s/m\rt)$ must be very small compared to its distance from the source.

In a recent paper \cite{Chicone1}, an approximation scheme has been introduced to deal with this extended pole-dipole system in
most astrophysical situations.
This scheme is based on the circumstance that $m^{-1} \, P^\mu - u_s^\mu$ is small and of order $\lt(M/r\rt)\lt[s/(mr)\rt]^2 \ll 1$.
In this approach, it then turns out that to first order in $s/(mr)$ one can assume that $P^\mu$ and $u_s^\mu$ are parallel, namely,
$P^\mu \approx m \, u_s^\mu$;
therefore, it follows from Eqs.~(\ref{Dixon-p})--(\ref{spin-condition}) in this case that
\be
{Du_s^\mu \over d\sg} & \approx & - {1 \over 2m} \, R^\mu{}_{\nu \alpha \beta} \, u_s^\nu \, S^{\alpha \beta}, \qquad
{DS^{\mu \nu} \over d\sg} \ \approx \ 0, \qquad S_{\mu \nu} \, u_s^\nu \ \approx \ 0.
\label{Dixon-approx}
\ee
The second section of Ref.~\cite{Chicone1} should be consulted for the details of the derivation of these equations.

In the absence of spin, system (\ref{Dixon-approx}) simply reduces to the geodesic equation
for the motion of the center of mass of the spinless extended particle, as expected~\cite{Ehlers}.
In Sec.~\ref{Sec2}, we solve system (\ref{Dixon-approx}) assuming that in the zeroth order of approximation the worldline
for a spinless particle is a timelike geodesic.
The motion away from the geodesic is then of first order in $s/(mr)$ in our approach.
This method is employed to study approximately circular motion near the equatorial plane of a Kerr source
of mass $M$ and angular momentum $J = Ma$ in Sec.~\ref{Sec3}.
That is, we assume that in the absence of spin, the test mass follows a stable equatorial circular geodesic orbit
in the exterior Kerr spacetime.
Our approximate analytic results based on system (\ref{Dixon-approx}) are compared with the numerical solution
of the extended pole-dipole system, Eqs.~(\ref{Dixon-p})--(\ref{spin-condition}), in Kerr spacetime in Sec.~\ref{Sec4}.
Section~\ref{Sec5} is devoted to the gravitomagnetic clock effect.
A brief discussion of our results is contained in Sec.~\ref{Sec6}.
The appendices contain further developments of our main results.

\section{First-Order Approximation}
\label{Sec2}

Consider the motion of an extended spinning test particle in a gravitational field.
Let $u^\mu$ be the four-velocity of the particle in the absence of spin and $\lambda^\mu{}_{\Cal}$ be the
orthonormal tetrad frame that is parallel propagated along the reference geodesic $x^\mu(\tau)$, where $u^\mu = dx^\mu/d\tau$
and $\tau$ is the proper time of the spinless particle.
More specifically, we imagine that along the unperturbed reference geodesic worldline $x^\mu (\tau)$, the Mathisson-Papapetrou
spin-curvature interaction is ``turned on'' at $\tau = 0$ and the spinning particle then follows $x_s^\mu (\tau)$ for $\tau > 0$.
Let $\delta x^\mu (\tau) = x_s^\mu (\tau) - x^\mu (\tau)$ be the deviation between the neighboring worldlines at the same time
$\tau > 0$ such that $\lt(\delta x^\mu\rt) u_\mu = 0$.
We will work to linear order in the small quantity $s/\lt(m \, r \rt) \ll 1$.
Here $\dl x^\mu(\tau)$ is a vector field defined orthogonally along $x^\mu(\tau)$ such that it connects the reference geodesic to the path of the
spinning particle in a way that amounts to a unique identification of events along the perturbed trajectory using the proper time of the
geodesic worldline $x^\mu$.
It follows that we can write
\be
\dl x^\mu & = & X^i(\tau) \, \lm^\mu{}_{\Ci},
\label{delta-x-mu}
\ee
where $\vec{X}(\tau)$ can be determined by the equations of motion (\ref{Dixon-approx}).
To this end, we establish a Fermi normal coordinate system along $x^\mu(\tau)$ based on the local frame $\lm^\mu{}_{\Cal}$ and solve
the equations of motion in this Fermi coordinate system.
The procedure that we follow is described in detail in Appendix~\ref{appendix:Fermi}.
Here we provide a general summary of our approximation method.

To linear order in our approximation scheme, it follows from Eq. (\ref{Dixon-approx}) that the spin tensor $S^{\mu \nu}$ is parallel propagated along the
unperturbed trajectory and $S_{\mu \nu} \, u^\nu = 0$.
One can then write
\be
S^{\mu \nu} & = & \lm^\mu{}_{\Ci} \, \lm^\nu{}_{\Cj} \, S^{\Ci \Cj},
\label{S-mu-nu}
\ee
where $S^{\Ci \Cj}$ are constants of the motion.
Moreover, it is simple to show that in the linear approximation scheme under consideration in this paper, $S_\mu \, u^\mu = 0$, so that
for $S_{\Cal} = S_\mu \, \lm^\mu{}_{\Cal}$ we have $S_{\ze} = 0$ and
\be
S_{\Ci} & = & {1 \over 2} \, \epsilon_{ijk} \, S^{\Cj \Ck}.
\label{S-vec-t}
\ee
Here, we have used the fact that $\epsilon_{\mu \nu \rho \sg} \lm^\mu{}_{\ze} \lm^\nu{}_{\on} \lm^\rho{}_{\tw} \lm^\sg{}_{\tr} = 1$,
since $\epsilon_{\ze \on \tw \tr} = 1$; hence, $\det \lt(\lm^\mu{}_{\Cal}\rt) = \lt(-g\rt)^{-1/2}$,
%
which is a consequence of the orthonormality of the tetrad frame.

Consider next the components of the spacetime curvature tensor measured by a free observer moving along the reference geodesic.
These components, $R_{\mu \nu \rho \sg} \, \lm^\mu{}_{\Cal} \lm^\nu{}_{\Cbt} \lm^\rho{}_{\Cgm} \lm^\sg{}_{\Cdl}$, can be expressed
as a $6 \times 6$ matrix $\lt({\cal R}_{\Dl \Sg}\rt)$, where the indices $\Dl$ and $\Sg$ range over the set \{01, 02, 03, 23, 31, 12\}.
One can then write
\be
{\cal R} \ = \
\left(%
\begin{array}{cc}
  {\cal E} & {\cal H} \\
  {\cal H}^\dag & {\cal D} \\
\end{array}%
\right),
\label{R-matrix-gen}
\ee
where ${\cal E}$, ${\cal H}$ and ${\cal D}$ are $3 \times 3$ matrices representing respectively the electric, magnetic and
spatial components of the Riemannian curvature.
In general, ${\cal E}$ and ${\cal D}$ are symmetric and ${\cal H}$ is traceless.
In a background Ricci-flat spacetime, ${\cal D} = -{\cal E}$, ${\cal E}$ is traceless and ${\cal H}$ is symmetric $\lt({\cal H} = {\cal H}^\dag \rt)$.
In this case, the curvature is completely determined by its electric and magnetic components.
These are given in general by
\be
{\cal E}_{\Ci \Cj} & = & R_{\mu \nu \rho \sg} \, \lm^\mu{}_{\ze} \, \lm^\nu{}_{\Ci} \, \lm^\rho{}_{\ze} \, \lm^\sg{}_{\Cj},
\label{Electric-R}
\nl
{\cal H}_{\Ci \Cj} & = & {1 \over 2} \, \epsilon_j{}^{kl} \, R_{\mu \nu \rho \sg} \, \lm^\mu{}_{\ze} \, \lm^\nu{}_{\Ci} \, \lm^\rho{}_{\Ck} \, \lm^\sg{}_{\Cl},
\label{Magnetic-R}
\ee
and, as described below, play important roles in our approximation scheme.

It is shown in Appendix~\ref{appendix:Fermi} that the worldline of the spinning particle is given, in our linear approximation, by
\be
x_s^\mu(\tau) & = & x^\mu(\tau) + X^i(\tau) \, \lm^\mu{}_{\Ci},
\label{x-s=}
\ee
where $\vec{X}(\tau)$ is a solution of the system
\be
{dX^i \over d\tau} & = & V^i,
\label{dx/dt=}
\nl
{dV^i \over d\tau} + {\cal E}_{\Ci \Cj}(\tau) \, X^j & = & {1 \over m} \, {\cal H}_{\Ci \Cj}(\tau) \, S^{\Cj},
\label{dv/dt=}
\ee
with the boundary conditions that at $\tau = 0$, $\vec{X}(0) = 0$ and $\Vvec(0) = 0$.
The unique solution of this system is thus due to the spin-curvature force folded together with the tidal influence of the
background gravitational field.
In fact, system (\ref{dx/dt=}) and (\ref{dv/dt=}) can be solved with the method of variation of parameters \cite{Chicone2} once the fundamental solution of the
homogeneous part is available.
In the absence of spin, the system simply reduces to the Jacobi equation in the Fermi coordinate system; therefore, system
(\ref{dx/dt=}) and (\ref{dv/dt=}) can be solved once the Jacobi fields along the path of the reference geodesic are completely known.
The general solution of system (\ref{dx/dt=}) and (\ref{dv/dt=}) is discussed in Appendix~\ref{appendix:Fermi}.
In the following section, we apply this method to Kerr spacetime; that is, we solve Eqs.~(\ref{dx/dt=}) and (\ref{dv/dt=}) for the
case of a spinning test particle that follows an approximately circular equatorial orbit in the exterior Kerr field.
The Kerr field, which represents the exact gravitational field of a rotating black hole, has been chosen here
for the sake of simplicity; in principle, one could employ any other exact solution of the gravitational field equations
describing the exterior field of a rotating mass \cite{Quevedo}.

\section{Kerr Spacetime}
\label{Sec3}

Consider circular geodesic orbits about a Kerr source of mass $M$ and angular momentum $J$.
In standard Boyer-Lindquist co-ordinates $(t, r, \th, \ph)$,
the geodesic equation for a circular equatorial orbit of fixed ``radius'' $r > 2M$ and
$\th = \pi/2$ reduces to
\be
t & = & {1 \over N} \lt(1 + a \, \Om_{\rm K}\rt) \tau, \qquad \ph \ = \ {1 \over N} \, \Om_{\rm K} \, \tau,
\label{geodesics-t-ph}
\ee
where $a = J/M > 0$ and we have chosen boundary conditions such that $t = \ph = 0$ at $\tau = 0$,
\be
N = \sqrt{1 - {3M \over r} + 2a \, \OmK}
\label{N}
\ee
and
\be
\OmK & = & \pm \sqrt{{M \over r^3}}
\label{Om-Kepler}
\ee
is the Keplerian frequency of the orbit.
Here the upper sign indicates an orbit that rotates in the same sense as the source,
while the lower sign indicates a counter-rotating orbit.
The energy $E$ and orbital angular momentum $L$ associated with these circular orbits are
given by
\be
E & = & {1 \over N} \lt(1 - {2M \over r} + a \, \OmK\rt), \qquad
L \ = \ {r^2 \OmK \over N} \lt(1 - 2a \, \OmK + {a^2 \over r^2}\rt).
\label{E-L}
\ee

%
A detailed discussion of Kerr geodesics can be found in Refs.~\cite{MTW,Chandra} and references therein.
Furthermore, the motion of spinning test particles in Kerr and other astrophysically interesting spacetimes has been studied
by a number of authors using other approaches, as can be seen from a perusal of
Refs.~\cite{Mashhoon1,Tod,Schmutzer,Abramowicz,Plyatsko,Semarak,Faruque1,Bini1,Tanaka,Suzuki,scatter,Hartl}.

It is possible to set up an orthonormal parallel-propagated frame $\lm^\mu{}_{\Cal}$ along the geodesic worldline of the spinless test particle
such that $\lm^\mu{}_{\ze} = dx^\mu/d\tau$ is the particle's four-velocity vector and $\lm^\mu{}_{\Ci}$, $i = 1,2,3$, are unit gyro
axes that form the particle's local spatial frame.
It follows that in $(t, r, \th, \ph)$ co-ordinates \cite{Chicone3}
\be
\lm^\mu{}_{\hat{0}} & = & \lt({1 + a \, \OmK \over N}, 0, 0, {\OmK \over N} \rt),
\label{tetrad-0}
\nl
\lm^\mu{}_{\hat{1}} & = & \lt(-{L \over r \, A} \, \sin \Th, \, A \, \cos \Th, \, 0,
 \, -{E \over r \, A} \, \sin \Th \rt),
\label{tetrad-1}
\nl
\lm^\mu{}_{\hat{2}} & = & \lt(0, 0, {1 \over r}, 0\rt),
\label{tetrad-2}
\nl
\lm^\mu{}_{\hat{3}} & = & \lt({L \over r \, A} \, \cos \Th, \, A \, \sin \Th, \, 0,
 \, {E \over r \, A} \, \cos \Th \rt),
\label{tetrad-3}
\ee
where $\Th = \OmK \tau$ and $A$ is given by
\be
A & = & \sqrt{1 - {2M \over r} + {a^2 \over r^2}}.
\label{A}
\ee
Furthermore, the electric and magnetic components of the spacetime curvature are given by \cite{Chicone3}
\be
{\cal E} \ = \
\left(%
\begin{array}{ccc}
  k_1 & 0 & k' \\
  0 & k_2 & 0 \\
  k' & 0 & k_3 \\
\end{array}%
\right),
\qquad
{\cal H} \ = \
\left(%
\begin{array}{ccc}
  0 & h & 0 \\
  h & 0 & h' \\
  0 & h' & 0 \\
\end{array}%
\right),
\label{E-H-matrix}
\ee
where $k_2 = -\lt(k_1 + k_3\rt)$ is a constant given by $k_2 = k\lt(3 \gm^2 -  2\rt)$ and
\be
k_1 & = & k\lt(1 - 3 \gm^2 \, \cos^2 \Th\rt), \quad k_3 \ = \ k\lt(1 - 3 \gm^2 \, \sin^2 \Th\rt), \quad k' \ = \ -3\gm^2 \, k \, \sin \Th \, \cos \Th,
\label{k}
\nl
h & = & -3\gm^2 \, \bt \, k \, \cos \Th, \qquad \hspace{3mm} h' \ = \ -3\gm^2 \, \bt \, k \, \sin \Th.
\label{h}
\ee
Here
\be
k & = & {M \over r^3} \ = \ \OmK^2,
\label{k0}
\ee
while $\bt$ and $\gm$ constitute a Lorentz pair, $\gm = 1/\sqrt{1 - \bt^2}$, given by
\be
\bt & = & {1 \over A} \lt(r \, \OmK - {a \over r}\rt), \qquad \gm \ = \ {A \over N}.
\label{bt-gm1}
\ee
%
The results presented here for the electric components of the curvature tensor are consistent with
the work of Marck \cite{Marck}.

We can now employ these results in order to find the perturbed orbit $x_s^\mu (\tau) = \lt(t_s, r_s, \th_s, \ph_s \rt)$
given by Eq. (\ref{x-s=}), namely,
\be
t_s & = & {1 + a \OmK \over N} \, \tau + {L \over rA} \lt(-X \, \sin \Th + Z \, \cos \Th \rt),
\label{ts=}
\nl
r_s & = & r + A\lt(X \, \cos \Th + Z \, \sin \Th \rt),
\label{rs=}
\nl
\th_s & = & {\pi \over 2} + {Y \over r},
\label{ths=}
\nl
\ph_s & = & {\OmK \over N} \, \tau + {E \over rA} \, \lt(-X \, \sin \Th + Z \, \cos \Th\rt),
\label{phs=}
\ee
where $\vec{X} = \lt(X, Y, Z\rt)$ and the tetrad frame given by Eqs.~(\ref{tetrad-0})--(\ref{tetrad-3}) has been used.
It is useful to write $S_{\Ci}$ in terms of spherical polar co-ordinates $(s, \vth, \vph)$ with respect to the local
tetrad frame as
\be
S_{\on} & = & s \, \sin \vth \, \cos \, \vph, \qquad S_{\tw} \ = \ - s \, \cos \vth, \qquad S_{\tr} \ = \ s \, \sin \vth \, \sin \, \vph,
\label{S-vec-comp}
\ee
so that $s \, \cos \vth$ is the component of the spin vector along the rotation axis of the Kerr source.
Then system (\ref{dx/dt=}) and (\ref{dv/dt=}) takes the form
\be
{d^2 X \over d \Th^2} & = & \lt(3 \gm^2 \cos^2 \Th -1 \rt)X + 3\gm^2 \, \sin \Th \, \cos \Th \, Z + 3 \gm^2 \, \bt \lt({s \over m} \, \cos \vth \rt) \cos \Th,
\label{d2X=}
\nl
{d^2 Y \over d \Th^2} & = & \lt(2 - 3 \gm^2\rt)Y - 3 \gm^2 \, \bt \lt({s \over m} \, \sin \vth \rt) \cos \lt(\Th - \vph\rt),
\label{d2Y=}
\nl
{d^2 Z \over d \Th^2} & = & 3 \gm^2 \sin \Th \, \cos \Th \, X + \lt(3\gm^2 \, \sin^2 \Th - 1\rt) Z + 3 \gm^2 \, \bt \lt({s \over m} \, \cos \vth \rt) \sin \Th,
\label{d2Z=}
\ee
with the boundary conditions that $\Xvec = \vec{0}$ and $\dot{\Xvec} = \vec{0}$ at $\Th = 0$.
Here an overdot denotes differentiation with respect to $\Th$.
Let us now consider a rotation about the vertical axis by an angle $\Th$ so that the rotated coordinate axes correspond to the radial,
vertical and tangential directions; that is, $Y' = Y$ and
\be
X' & = & X \, \cos \Th + Z \, \sin \Th, \qquad Z' \ = \ - X \, \sin \Th + Z \, \cos \Th.
\label{X'Z'=}
\ee
We note that $x_s^\mu$ can be conveniently expressed in terms of the rotated coordinates $\lt(X',Y',Z'\rt)$, as they appear explicitly in
Eqs.~(\ref{ts=})--(\ref{phs=}) for the perturbed orbit.

It is possible to show that under rotation (\ref{X'Z'=}), Eqs.~(\ref{d2X=})--(\ref{d2Z=}) take the form
\be
\ddot{X}' - 2 \, \dot{Z}' - 3 \gm^2 \, X' & = & 3 \gm^2 \, \bt \lt({s \over m} \, \cos \vth\rt),
\label{d2X-rot=}
\nl
\ddot{Y}' + \lt(3\gm^2 - 2\rt) Y' & = & -3 \gm^2 \, \bt \lt({s \over m} \, \sin \vth\rt) \cos \lt(\Th - \vph\rt),
\label{d2Y-rot=}
\nl
\ddot{Z}' + 2 \, \dot{X}' & = & 0,
\label{d2Z-rot=}
\ee
with the boundary conditions that at $\Th = 0$, $\Xvec' = \vec{0}$ and $\dot{\Xvec}' = \vec{0}$.
The unique solution of this system is given by
\be
X' & = & {3 \over \rho^2} \, \gm^2 \, \bt \lt({s \over m} \cos \vth\rt) \lt[1 - \cos \lt(\rho \, \Th\rt)\rt],
\label{X'-sol=}
\nl
Y' & = & {1 \over \bt} \lt({s \over m} \sin \vth\rt) \lt[\cos \vph \, \cos \lt(\zeta \, \Th\rt) + {1 \over \zeta} \, \sin \vph \, \sin \lt(\zeta \, \Th\rt)
- \cos \lt(\Th - \vph\rt)\rt],
\label{Y'-sol=}
\nl
Z' & = & -{6 \over \rho^3} \, \gm^2 \, \bt \lt({s \over m} \cos \vth\rt) \lt[\rho \, \Th - \sin \lt(\rho \, \Th\rt) \rt].
\label{Z'-sol=}
\ee
Here
\be
\rho & = & \sqrt{4 - 3 \gm^2}, \qquad \zeta \ = \ \sqrt{3 \gm^2 - 2}
\label{lm,sg=}
\ee
are such that $\rho |\OmK|$ and $\zeta |\OmK|$ are the proper radial and vertical epicyclic frequencies, respectively.
%

It follows from these results that the orbit of the spinning particle in the standard Kerr coordinate system is given by
\be
t_{\rm s} & = & {1 + a \, \OmK \over N} \, \tau + {6 \over \rho^3} \, \gm^2 \, \bt \lt({s \over m \, r} \, \cos \vth\rt) {L \over A}
\lt[\sin \lt(\rho \, \Th\rt) - \rho \, \Th \rt],
\label{t-s}
\nl
r_{\rm s} & = & r \lt\{1 + {3 \over \rho^2} \, \gm^2 \, \bt \lt({s \over m \, r} \, \cos \vth\rt) A \lt[1 - \cos \lt(\rho \, \Th\rt)\rt] \rt\},
\label{r-s}
\nl
\th_{\rm s} & = & {\pi \over 2} + {1 \over \bt} \lt({s \over m \, r} \, \sin \vth\rt)
\lt[\cos \vph \, \cos \lt(\zeta \, \Th\rt) + {1 \over \zeta} \, \sin \vph \, \sin \lt(\zeta \, \Th\rt) - \cos \lt(\Th - \vph\rt) \rt],
\label{th-s}
\nl
\ph_{\rm s} & = & {\OmK \over N} \, \tau + {6 \over \rho^3} \, \gm^2 \, \bt \lt({s \over m \, r} \, \cos \vth\rt) {E \over A} \,
\lt[\sin \lt(\rho \, \Th\rt) - \rho \, \Th\rt].
\label{ph-s}
\ee
The perturbation in $\ph_s$ is simply $E/L$ times the perturbation in $t_s$.
In Eqs.~(\ref{t-s})--(\ref{ph-s}), the temporal parameter $\tau$ can be replaced by $\sg$,
which is the proper time of the perturbed worldline, since $\tau = \sg + O(s^2)$ by
Eqs.~(\ref{A7})--(\ref{A9}) of Appendix A.
It is interesting to note here the role of spin length scales
\be
s^* & = & \lt(s \over m\rt) \cos \vth, \qquad \tilde{s} \ = \ \lt(s \over m\rt) \sin \vth
\label{s-scales}
\ee
in the worldline of the particle; the perturbations in $t_s$, $r_s$, and $\phi_s$ are proportional to $s^*$, while the polar motion
away from the equatorial plane is proportional to $\tilde{s}$.

Equations~(\ref{t-s})--(\ref{ph-s}), involving the first-order perturbation terms proportional to $s/(mr)$, should be viewed as containing the zeroth and
first-order terms of a perturbation expansion in powers of a sufficiently small parameter $s/(mr) \ll 1$.
Moreover, for a black hole $\lt(a \leq M\rt)$, $\beta$ in Eqs.~(\ref{t-s})--(\ref{ph-s}) is positive for co-rotating orbits and negative for counter-rotating orbits,
while $\gamma^2$ monotonically decreases from $4/3$ at the last stable circular orbit to unity at infinity.
Similarly, $|\beta|$ decreases monotonically from $1/2$ to zero over the same range; in fact, for $r \gg M$ and $r \gg a$, $|\beta| \approx \sqrt{M/r}$.

It is interesting to mention here the behavior of $\bt$ for $a > M$.
As before, $\bt$ is negative for counter-rotating orbits, while for co-rotating orbits $\bt$ can be positive or negative.
In fact, for a co-rotating orbit in this case $(a > M)$, $\bt$ is positive for $r > a^2/M$, negative for $r < a^2/M$, and
zero for $r = a^2/M$.

In Eq.~(\ref{r-s}), it is important to note the periodic motion in the radius:
it oscillates between $r$ and $r + 6 \, A \, \gm^2 \, \bt \, s^*/\rho^2$ with a proper period of $2\pi/\lt(\rho |\OmK|\rt)$.
For the motion of the Earth about the Sun, $r \gg M$ and the M{\o}ller radius of the Earth is about 200 cm, so that
the amplitude of this periodic variation in the astronomical unit amounts to
\be
{\dl r \over r} & \approx & {6 \over c} \, \lt|\OmK\rt|s^* \ \sim \ 10^{-14},
\label{r-variation}
\ee
which is a few orders of magnitude too small to be measurable at present.
In this connection, it is interesting to note that a {\it secular} increase of the astronomical unit,
amounting to about $10^3$ cm per century, has been recently reported based on the analysis of radiometric data \cite{Standish,Lammerzahl}.
Furthermore, the polar motion about the equatorial plane as given by Eq.~(\ref{th-s}) consists of a superposition of
two frequencies:  $\lt|\OmK\rt|$ and $\zeta\lt|\OmK\rt|$.
For $r \gg M \geq a$, this motion in the polar direction exhibits a beat phenomenon.
That is, the ``fast'' harmonic oscillation at essentially the Keplerian frequency $\lt|\OmK\rt|$ is modulated by a
``slow'' beat frequency $ \approx \omega_g/2$, where $\omega_g = 3GM \lt|\OmK\rt|/\lt(2c^2 \, r\rt)$ is the
``geodetic'' (i.e., de~Sitter-Fokker) precession frequency for the unperturbed orbit.
The net amplitude (in radians) of the polar motion for $r \gg M \geq a$ is independent of $c$ and is given by
$\tilde{s}/\sqrt{GMr}$.
Nevertheless, this maximum amplitude is a relativistic effect and builds up over a ``long'' timescale;
this circumstance is reminiscent of another phenomenon related to relativistic nutation \cite{Mashhoon1a}.
For the Earth, the yearly spin-induced up and down polar motion about the ecliptic has a net amplitude of about
$10^6$ cm and beat period of $10^8$ years; that is, the amplitude of the polar motion away from the ecliptic develops
gradually and reaches a maximum of about 10 kilometers over a period of about 25 million years.

It would be very interesting to search for these radial and polar motions in the strong-field regime close
to black holes or in relativistic binary systems.

Let us now compare and contrast these analytic (but approximate) results with the numerical analysis of the exact equations in Sec.~IV.


\section{Numerical Integration of MPD Equations}
\label{Sec4}

To determine the domain of validity of our linear approximation scheme, it is interesting to make a numerical
comparison of our results with the full MPD equations, which must be written in a form that is suitable for numerical integration.
A direct comparison is possible once the reduced MPD equations are expressed in terms of proper time $\sg$ as \cite{Singh1}
\be
{dP^\alpha \over d\sg} & = & - \Gamma^\alpha{}_{\mu \nu} \, P^\mu \, u_s^\nu
\nn
& &{} + {1 \over 2m} \, R^\alpha{}_{\beta \rho \sigma} \,
\epsilon^{\rho \sigma}{}_{\mu \nu} \, S^\mu \, P^\nu \, u_s^\beta,
\label{dp/dt}
\nl
{dS^\alpha \over d\sg} & = & - \Gamma^\alpha{}_{\mu \nu} \, S^\mu \, u_s^\nu
\nn
& &{} + \left({1 \over 2m^3} \, R_{\gamma \beta \rho \sigma} \,
\epsilon^{\rho \sigma}{}_{\mu \nu} \, S^\mu \, P^\nu \, S^\gamma \, u_s^\beta \right)P^\alpha,
\label{ds/dt}
\nl
{dx^\alpha \over d\sg} & = & u_s^\al \ = \
{\cal N} \left(P^\alpha
+ {1 \over 2} \, {S^{\alpha \beta} \, R_{\beta \gamma \mu \nu} \, P^\gamma \, S^{\mu \nu}
\over m^2 + {1 \over 4} \, R_{\mu \nu \rho \sigma} \, S^{\mu \nu} \, S^{\rho \sigma}} \right) ,
\label{dx/dt}
\ee
where ${\cal N} = -u_s^\al \, P_\al/m^2$ is a normalization parameter such that $g_{\mu \nu} \, u_s^\mu \, u_s^\nu = -1$ throughout the spinning particle's motion.

\begin{figure*}
\psfrag{sg (M)}[cc][][2.5][0]{\hspace{0.5cm} $\sg$ [M]}
\psfrag{r (M)}[bc][][2.5][0]{\hspace{0.5cm} $r_s(\sg)$ [M]}
\psfrag{dr (M)}[bc][][2.5][0]{\hspace{0.5cm} $\Delta r_s(\tau)$ [M]}
\psfrag{theta (radians)}[bc][][2.5][0]{\hspace{0.5cm} $\th_s(\sg)$ [radians]}
\psfrag{dtheta (radians)}[bc][][2.5][0]{\hspace{0.5cm} $\Delta \th_s(\tau)$ [radians]}
\psfrag{phi (radians)}[bc][][2.5][0]{\hspace{0.5cm} $\ph_s(\sg)$ [radians]}
\psfrag{dphi (radians)}[bc][][2.5][0]{\hspace{0.5cm} $\Delta \ph_s(\tau)$ [radians]}
\begin{minipage}[t]{0.3 \textwidth}
\centering
\subfigure[\hspace{0.2cm} $s/(mr) = 10^{-3}$ (MPD)]{
\label{fig:r-MPD-s/(mr)=1e-3}
\rotatebox{0}{\includegraphics[width = 6.6cm, height = 5.0cm, scale = 1]{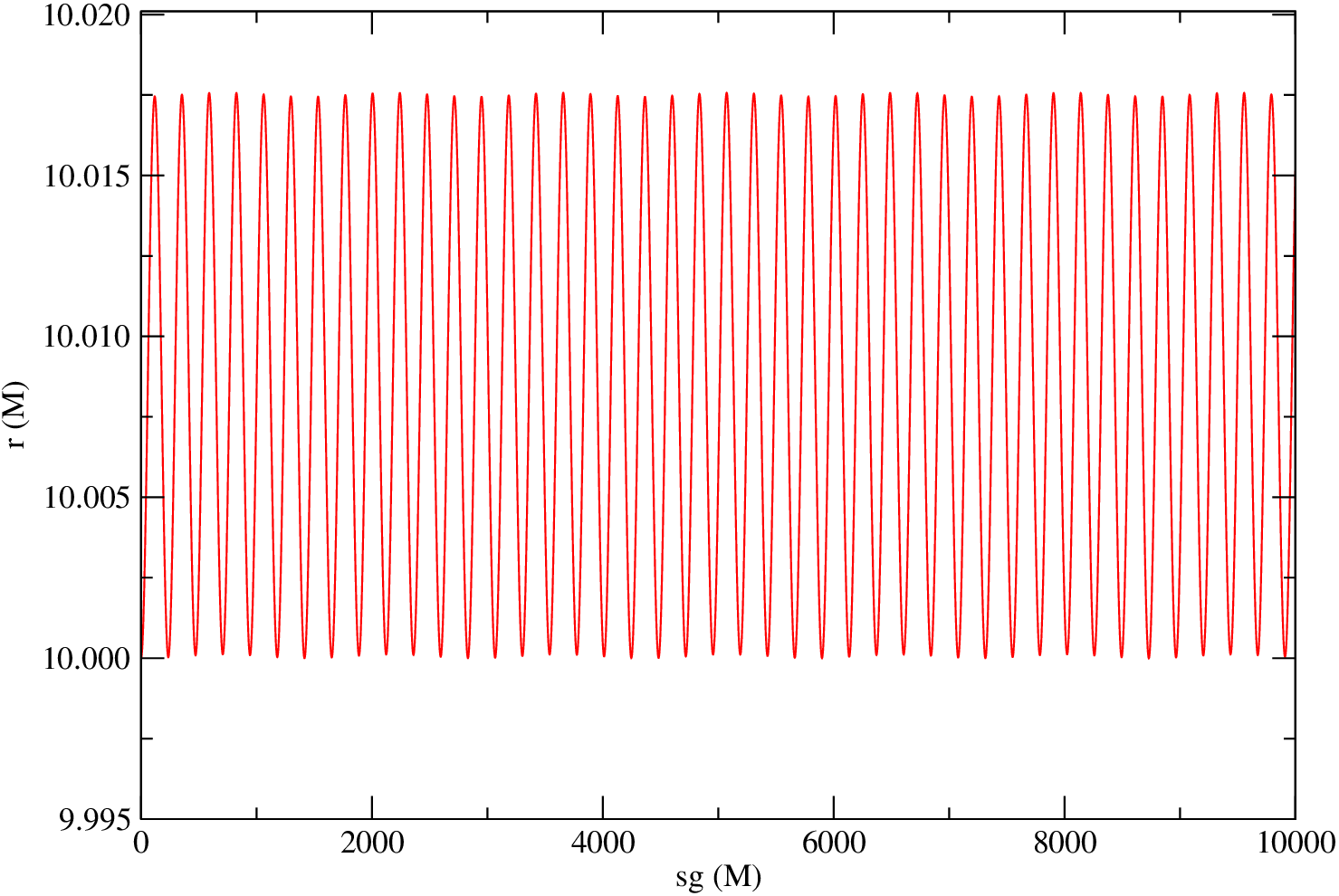}}}
\end{minipage}
\hspace{2.0cm}
\begin{minipage}[t]{0.3 \textwidth}
\centering
\subfigure[\hspace{0.2cm} $s/(mr) = 10^{-3}$ (Linear)]{
\label{fig:r-lin-s/(mr)=1e-3}
\rotatebox{0}{\includegraphics[width = 6.6cm, height = 5.0cm, scale = 1]{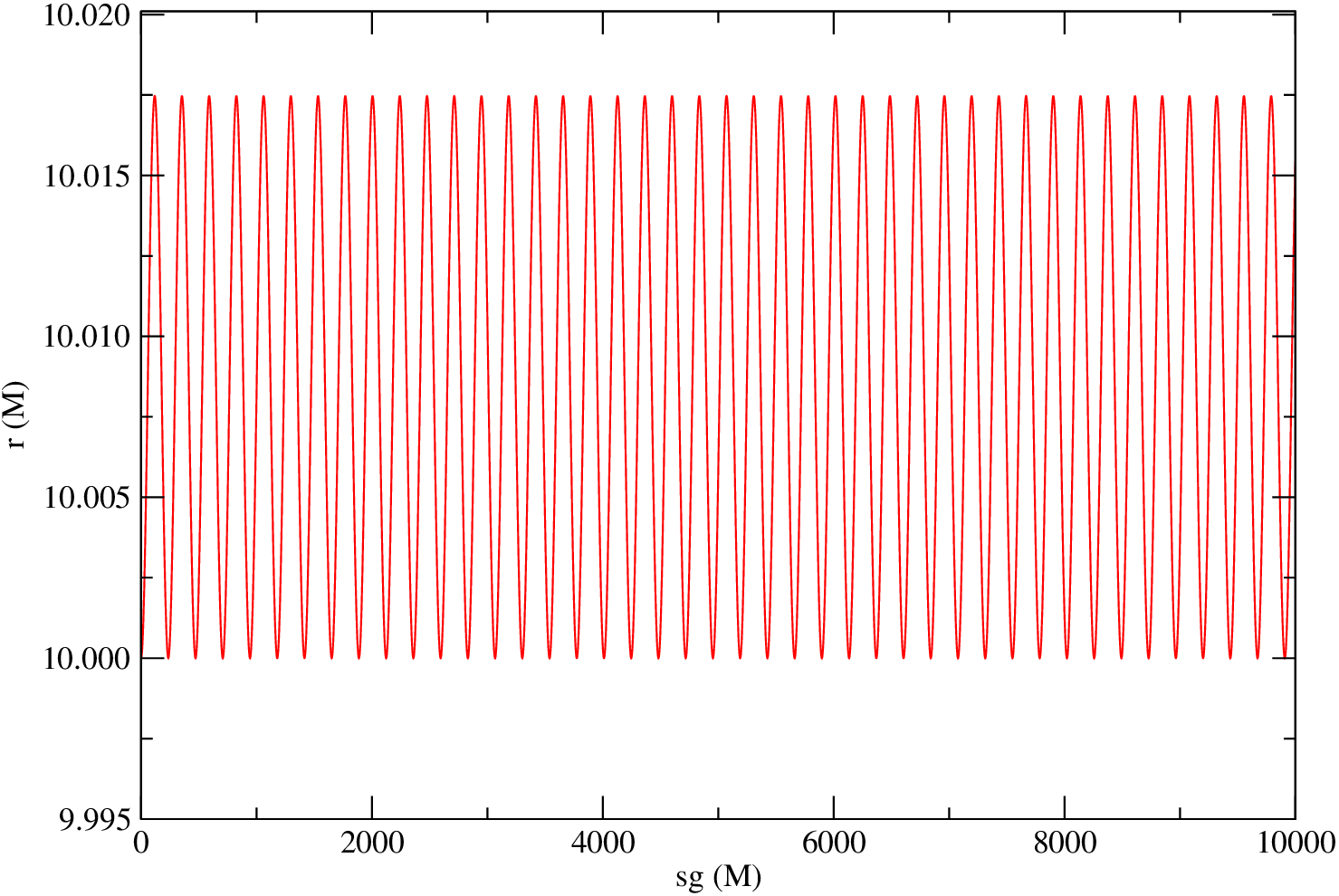}}}
\end{minipage} \\
\vspace{0.8cm}
\begin{minipage}[t]{0.3 \textwidth}
\centering
\subfigure[\hspace{0.2cm} $s/(mr) = 10^{-2}$ (MPD)]{
\label{fig:r-MPD-s/(mr)=1e-2}
\rotatebox{0}{\includegraphics[width = 6.6cm, height = 5.0cm, scale = 1]{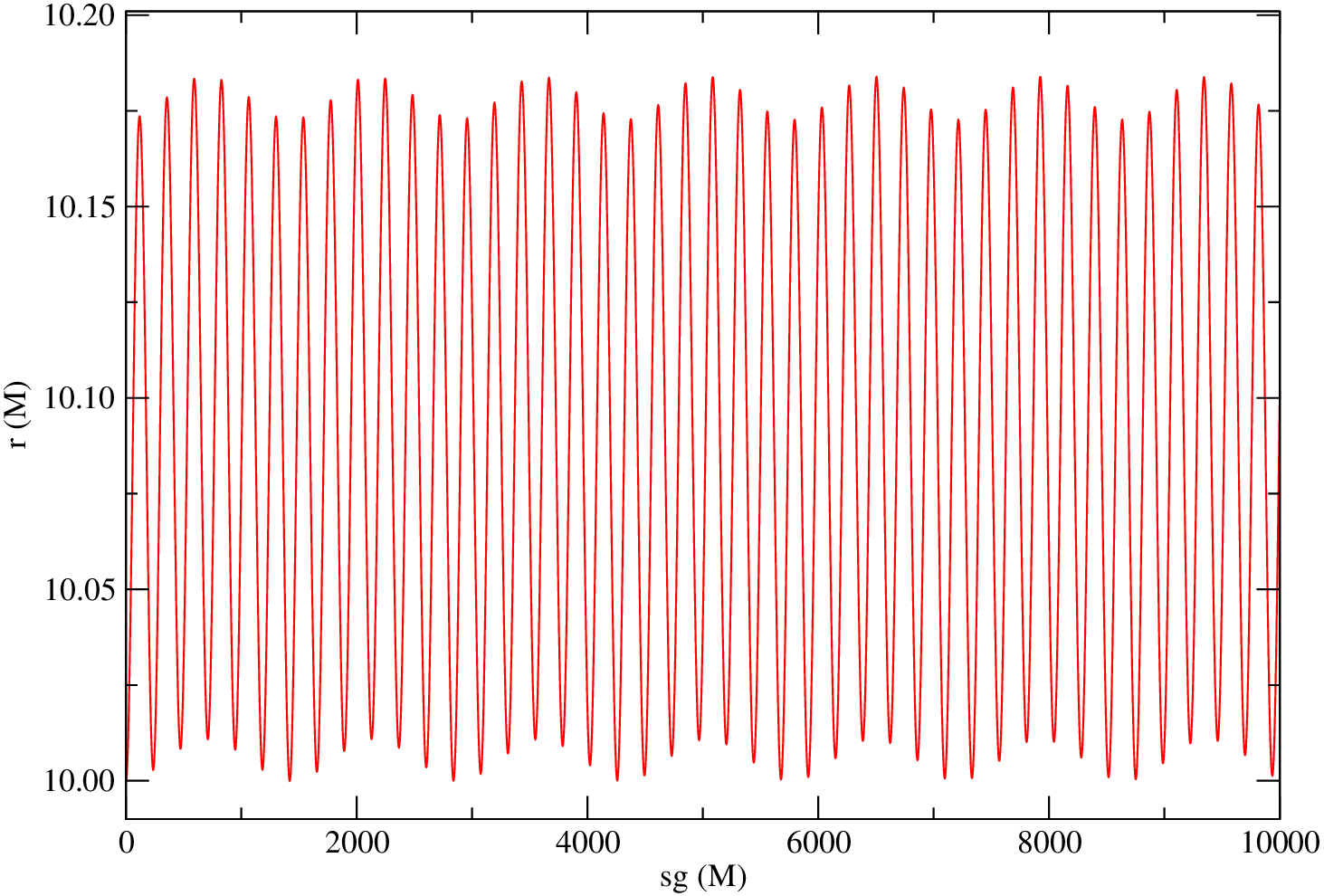}}}
\end{minipage}
\hspace{2.0cm}
\begin{minipage}[t]{0.3 \textwidth}
\centering
\subfigure[\hspace{0.2cm} $s/(mr) = 10^{-2}$ (Linear)]{
\label{fig:r-lin-s/(mr)=1e-2}
\rotatebox{0}{\includegraphics[width = 6.6cm, height = 5.0cm, scale = 1]{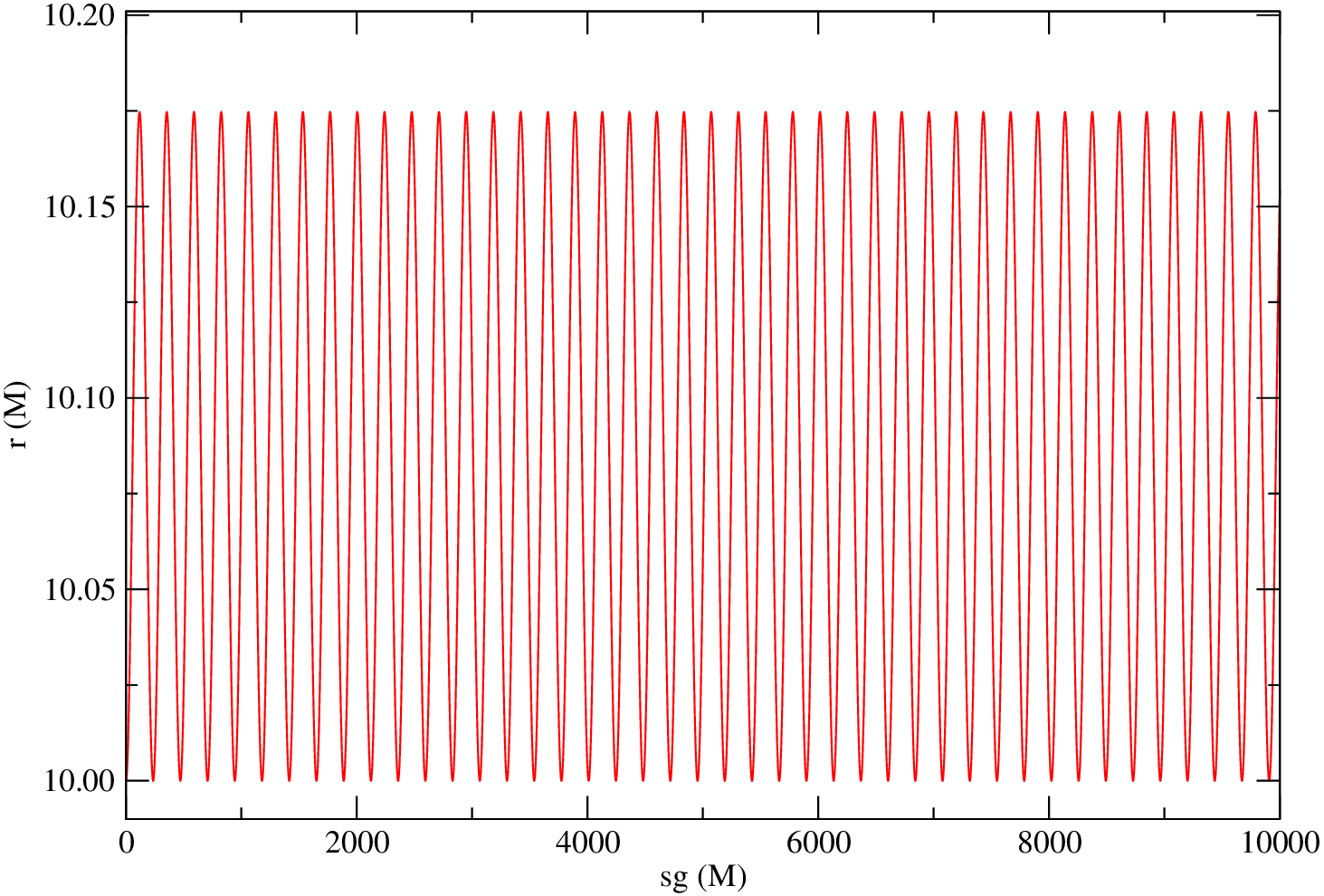}}}
\end{minipage} \\
\vspace{0.8cm}
\begin{minipage}[t]{0.3 \textwidth}
\centering
\subfigure[\hspace{0.2cm} $s/(mr) = 10^{-1}$ (MPD)]{
\label{fig:r-MPD-s/(mr)=1e-1}
\rotatebox{0}{\includegraphics[width = 6.6cm, height = 5.0cm, scale = 1]{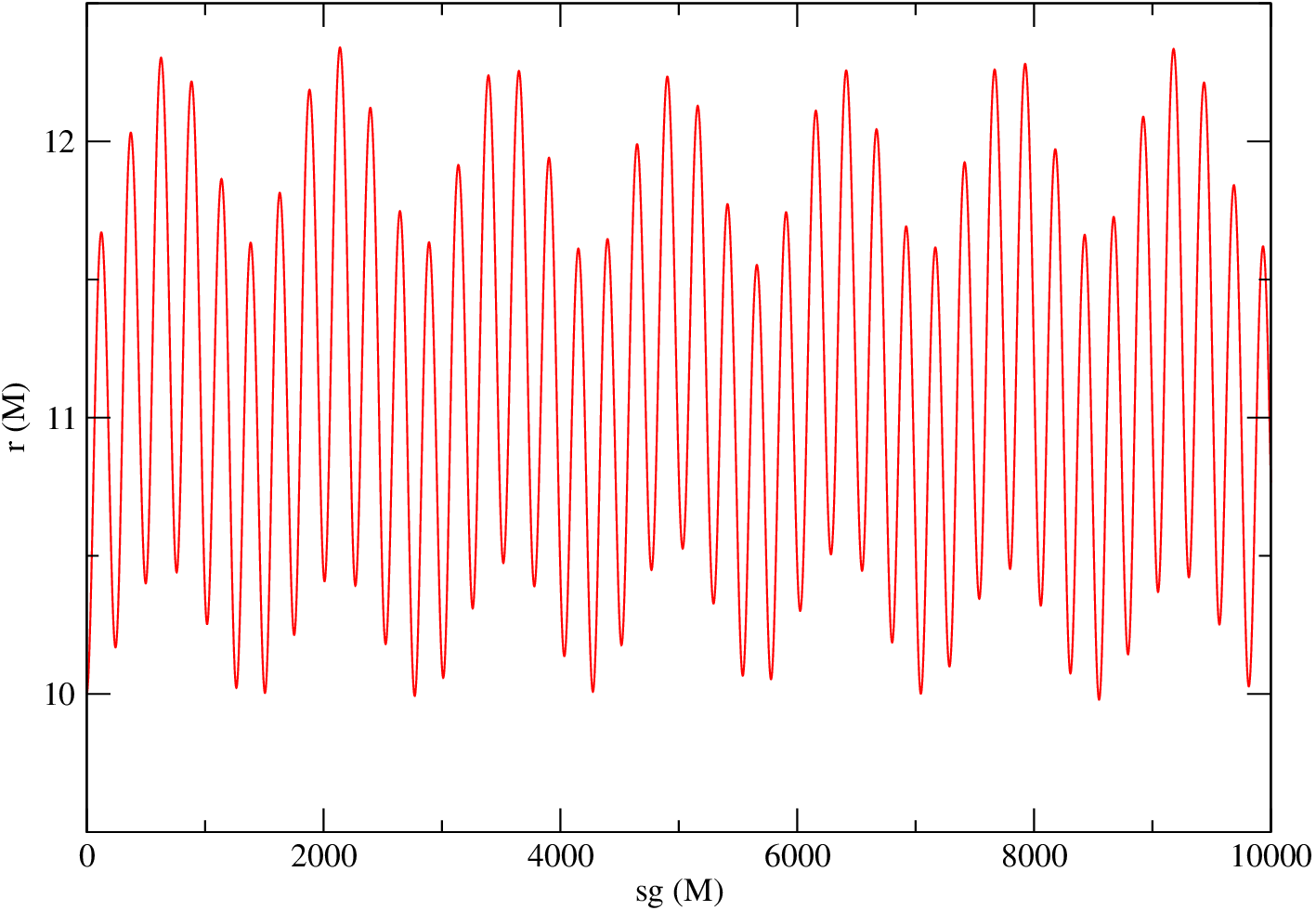}}}
\end{minipage}
\hspace{2.0cm}
\begin{minipage}[t]{0.3 \textwidth}
\centering
\subfigure[\hspace{0.2cm} $s/(mr) = 10^{-1}$ (Linear)]{
\label{fig:r-lin-s/(mr)=1e-1}
\rotatebox{0}{\includegraphics[width = 6.6cm, height = 5.0cm, scale = 1]{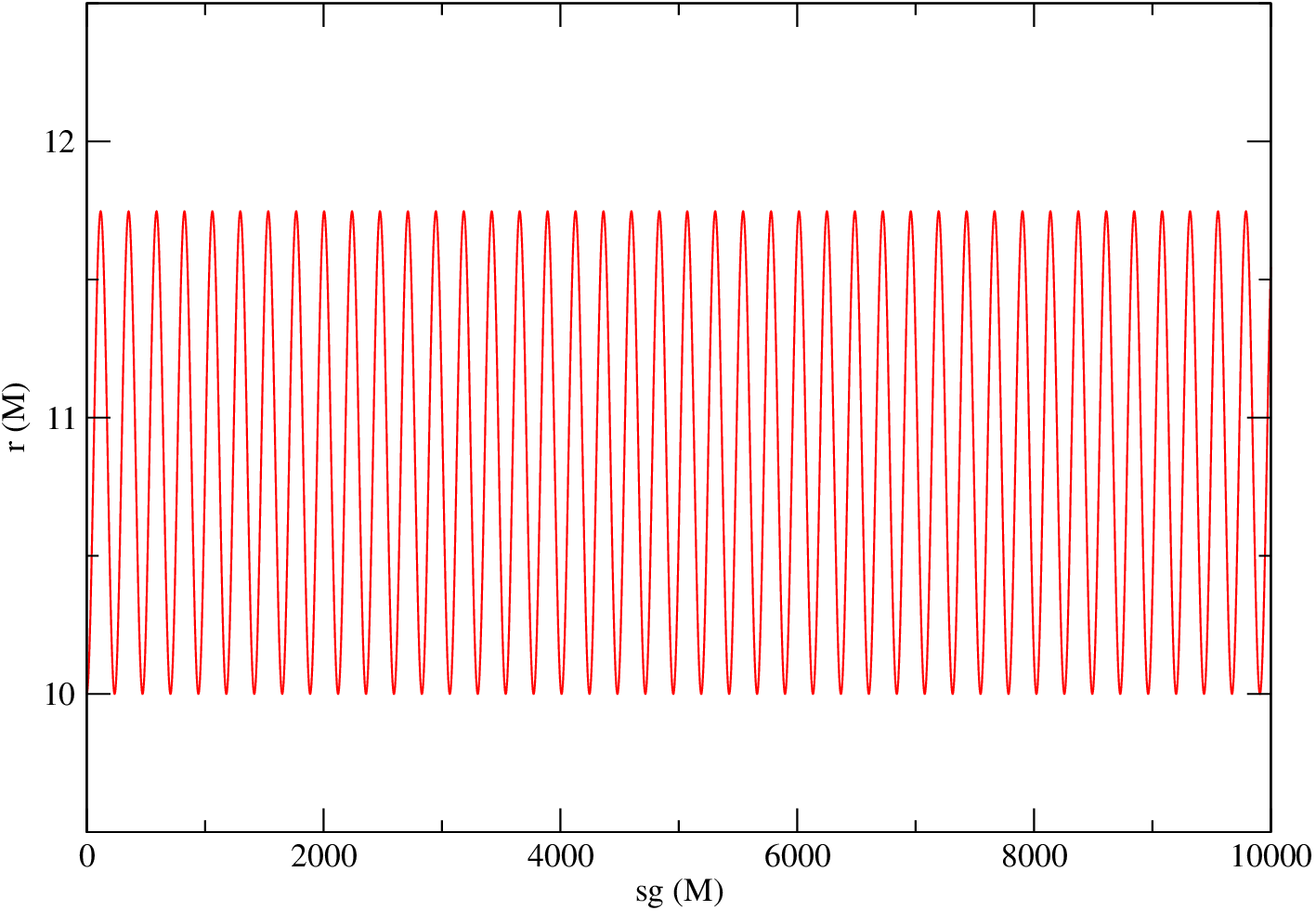}}}
\end{minipage} \\
\caption{\label{fig:r} The orbital radial co-ordinate $r_s(\sg)$ for various choices of $s/(mr)$,
where $m = 10^{-2} \, M$, $r = 10 \, M$, $\vth = \vph = \pi/4$, and $a = 0.50$.
}
\end{figure*}

\begin{figure*}
\psfrag{sg (M)}[cc][][2.5][0]{\hspace{0.5cm} $\sg$ [M]}
\psfrag{r (M)}[bc][][2.5][0]{\hspace{0.5cm} $r_s(\sg)$ [M]}
\psfrag{dr (M)}[bc][][2.5][0]{\hspace{0.5cm} $\Delta r_s(\tau)$ [M]}
\psfrag{theta (radians)}[bc][][2.5][0]{\hspace{0.5cm} $\th_s(\sg)$ [radians]}
\psfrag{dtheta (radians)}[bc][][2.5][0]{\hspace{0.5cm} $\Delta \th_s(\tau)$ [radians]}
\psfrag{phi (radians)}[bc][][2.5][0]{\hspace{0.5cm} $\ph_s(\sg)$ [radians]}
\psfrag{dphi (radians)}[bc][][2.5][0]{\hspace{0.5cm} $\Delta \ph_s(\tau)$ [radians]}
\begin{minipage}[t]{0.3 \textwidth}
\centering
\subfigure[\hspace{0.2cm} $s/(mr) = 10^{-3}$ (MPD)]{
\label{fig:theta-MPD-s/(mr)=1e-3}
\rotatebox{0}{\includegraphics[width = 6.6cm, height = 5.0cm, scale = 1]{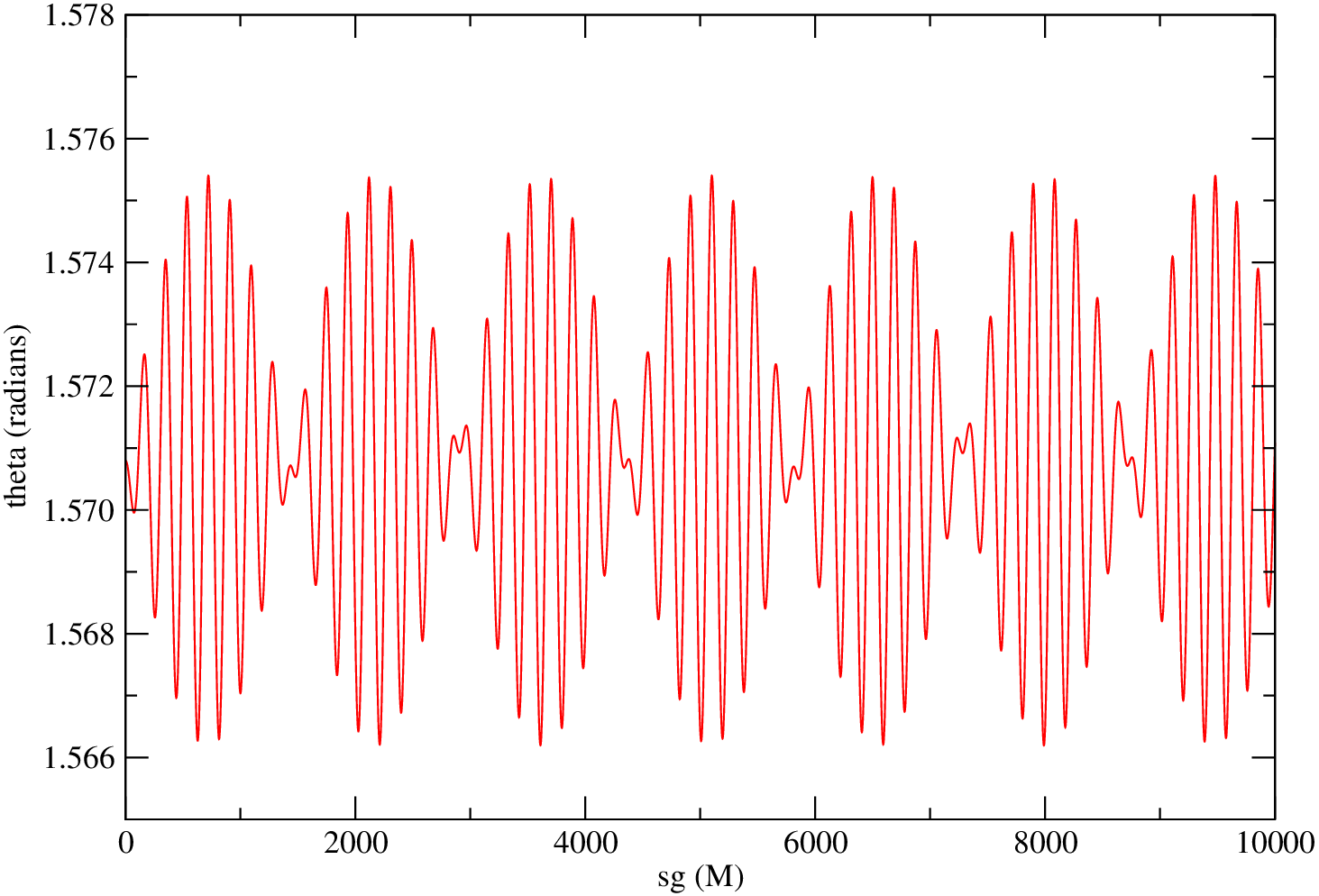}}}
\end{minipage}
\hspace{2.0cm}
\begin{minipage}[t]{0.3 \textwidth}
\centering
\subfigure[\hspace{0.2cm} $s/(mr) = 10^{-3}$ (Linear)]{
\label{fig:theta-lin-s/(mr)=1e-3}
\rotatebox{0}{\includegraphics[width = 6.6cm, height = 5.0cm, scale = 1]{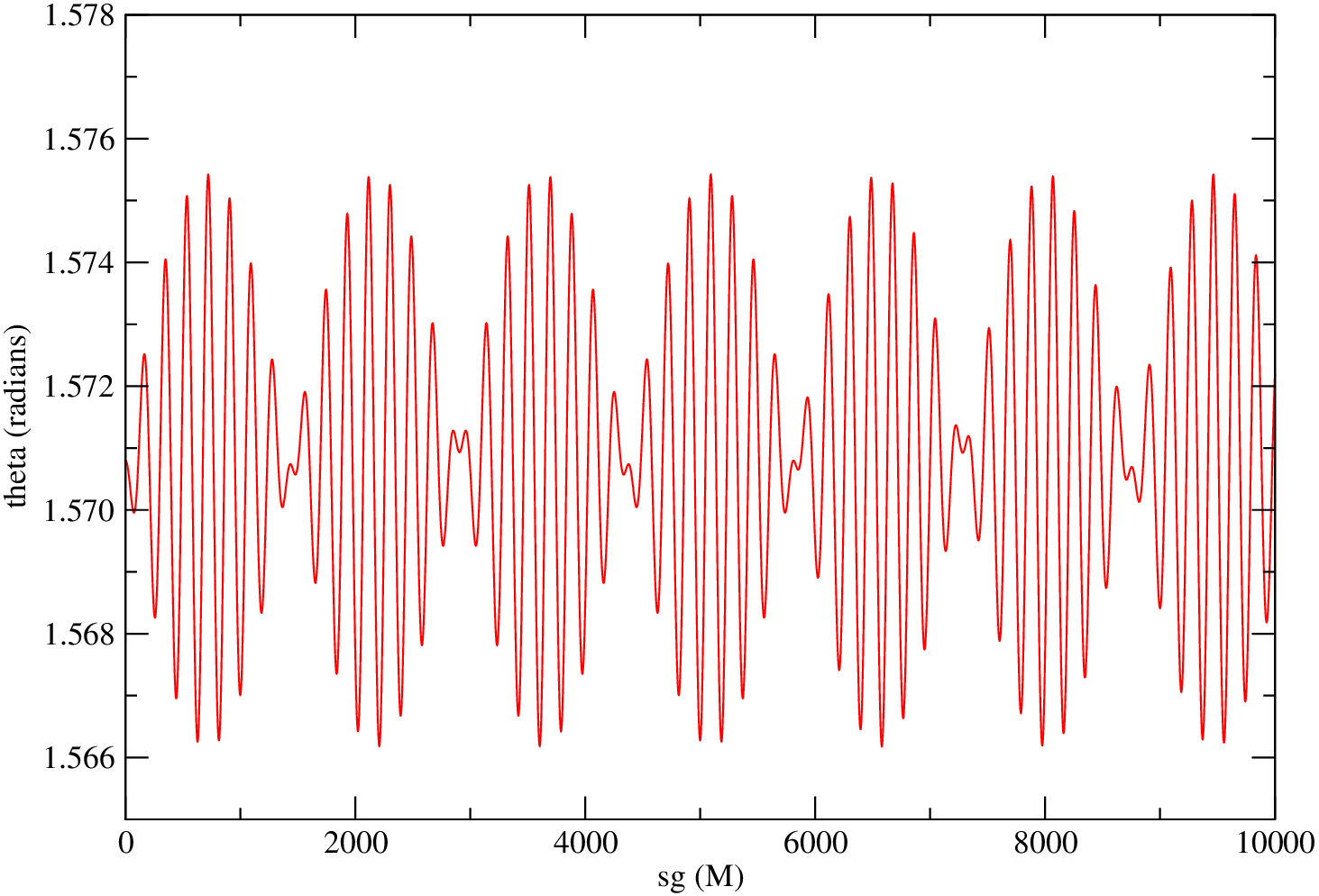}}}
\end{minipage} \\
\vspace{0.8cm}
\begin{minipage}[t]{0.3 \textwidth}
\centering
\subfigure[\hspace{0.2cm} $s/(mr) = 10^{-2}$ (MPD)]{
\label{fig:theta-MPD-s/(mr)=1e-2}
\rotatebox{0}{\includegraphics[width = 6.6cm, height = 5.0cm, scale = 1]{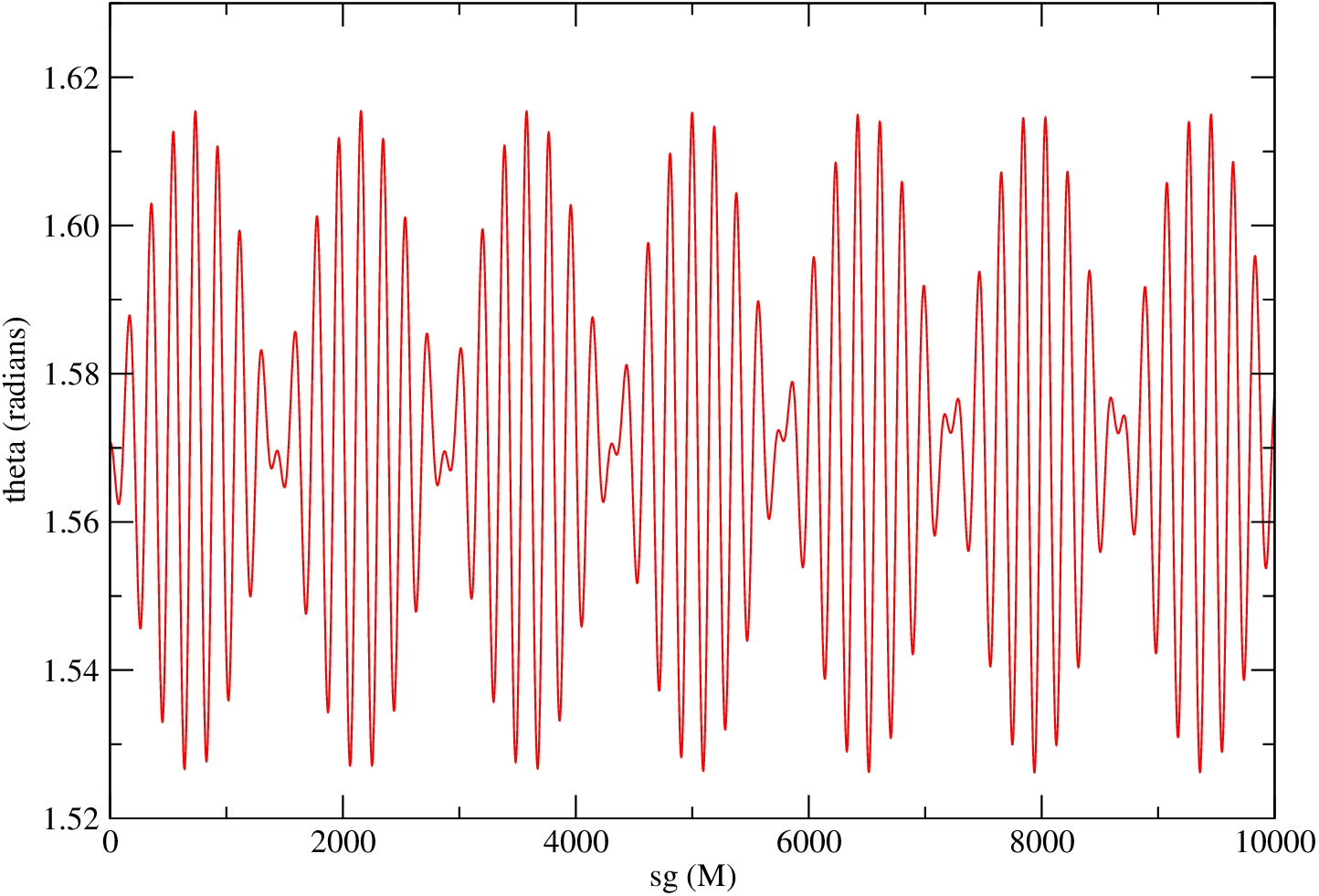}}}
\end{minipage}
\hspace{2.0cm}
\begin{minipage}[t]{0.3 \textwidth}
\centering
\subfigure[\hspace{0.2cm} $s/(mr) = 10^{-2}$ (Linear)]{
\label{fig:theta-lin-s/(mr)=1e-2}
\rotatebox{0}{\includegraphics[width = 6.6cm, height = 5.0cm, scale = 1]{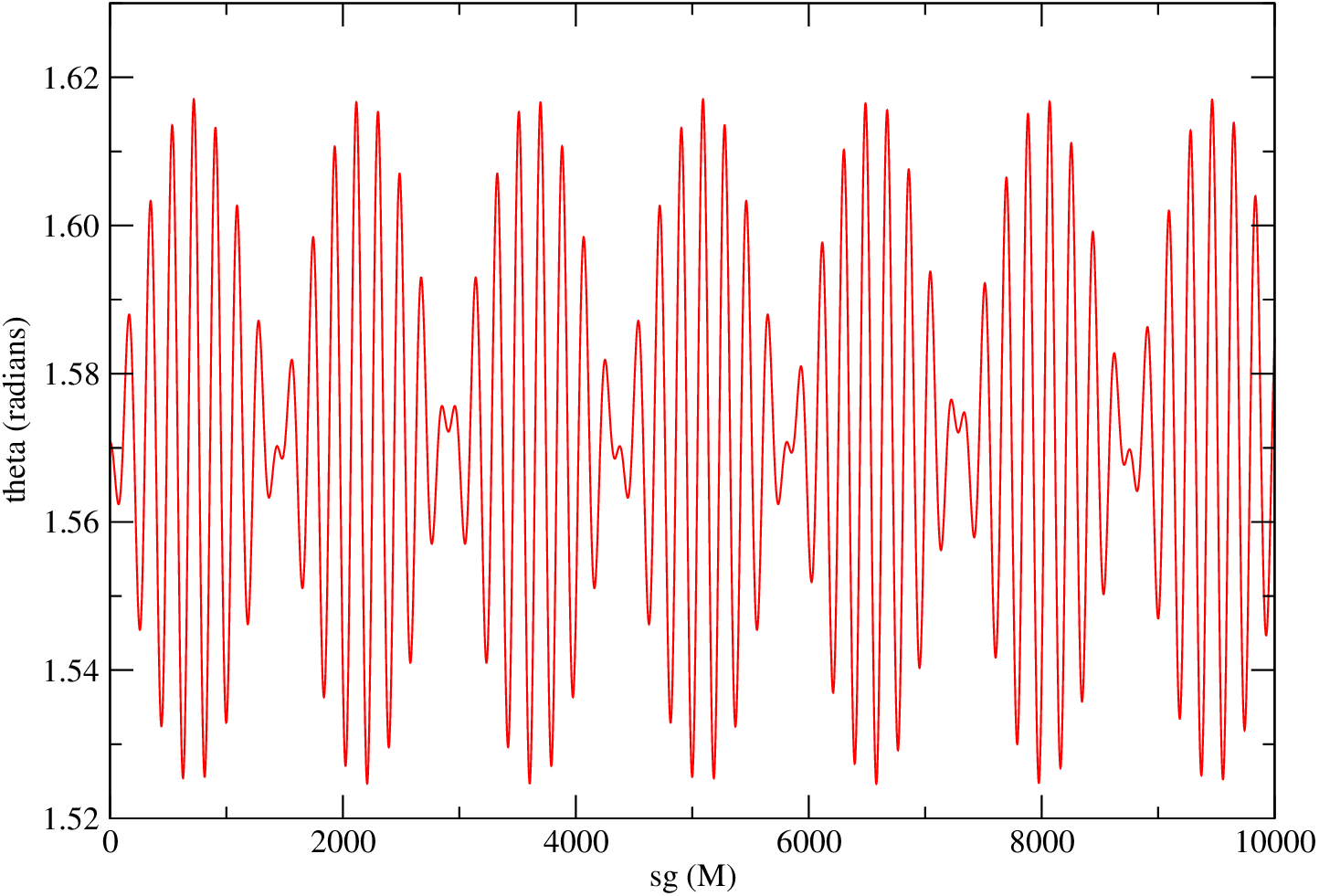}}}
\end{minipage} \\
\vspace{0.8cm}
\begin{minipage}[t]{0.3 \textwidth}
\centering
\subfigure[\hspace{0.2cm} $s/(mr) = 10^{-1}$ (MPD)]{
\label{fig:theta-MPD-s/(mr)=1e-1}
\rotatebox{0}{\includegraphics[width = 6.6cm, height = 5.0cm, scale = 1]{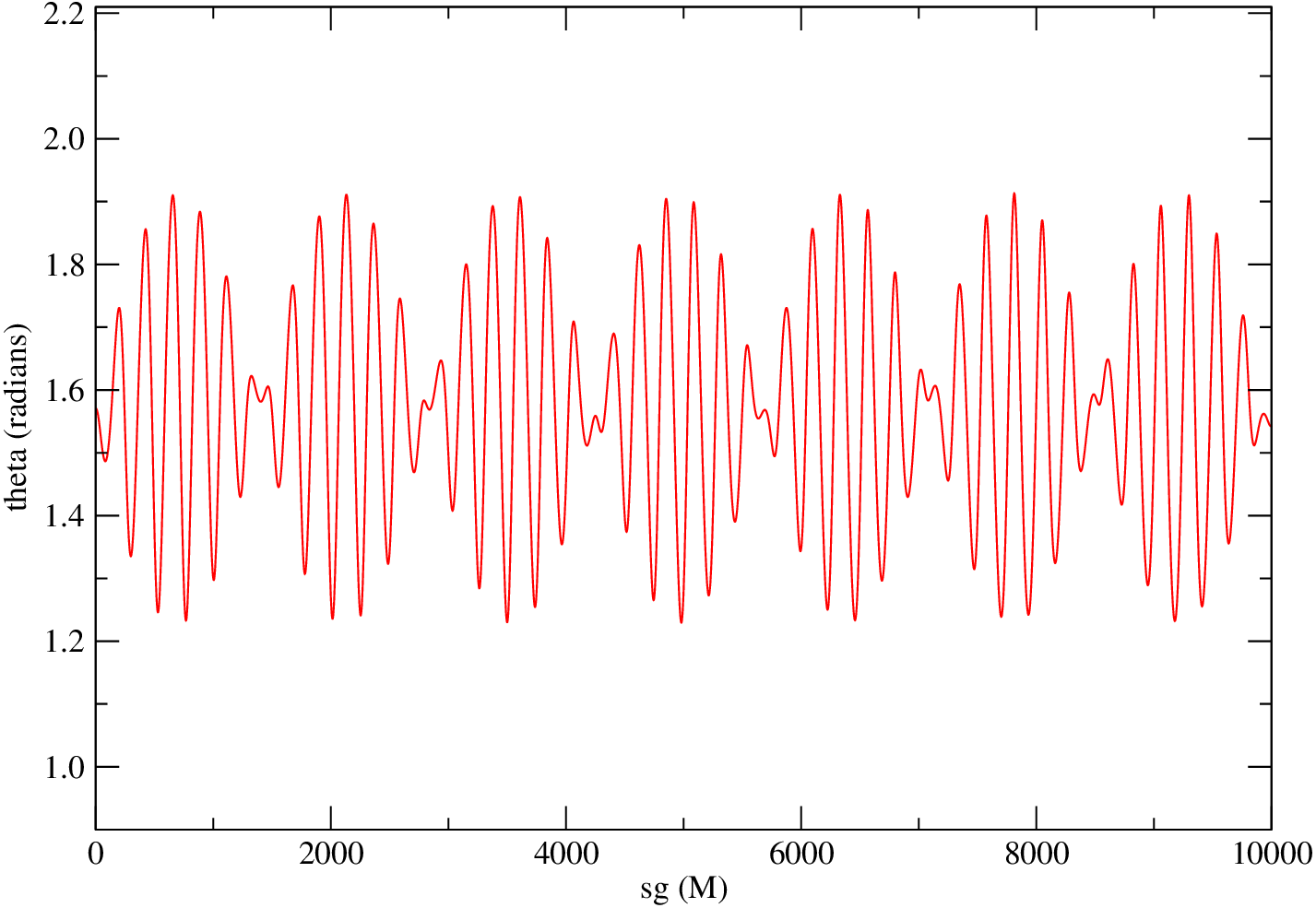}}}
\end{minipage}
\hspace{2.0cm}
\begin{minipage}[t]{0.3 \textwidth}
\centering
\subfigure[\hspace{0.2cm} $s/(mr) = 10^{-1}$ (Linear)]{
\label{fig:theta-lin-s/(mr)=1e-1}
\rotatebox{0}{\includegraphics[width = 6.6cm, height = 5.0cm, scale = 1]{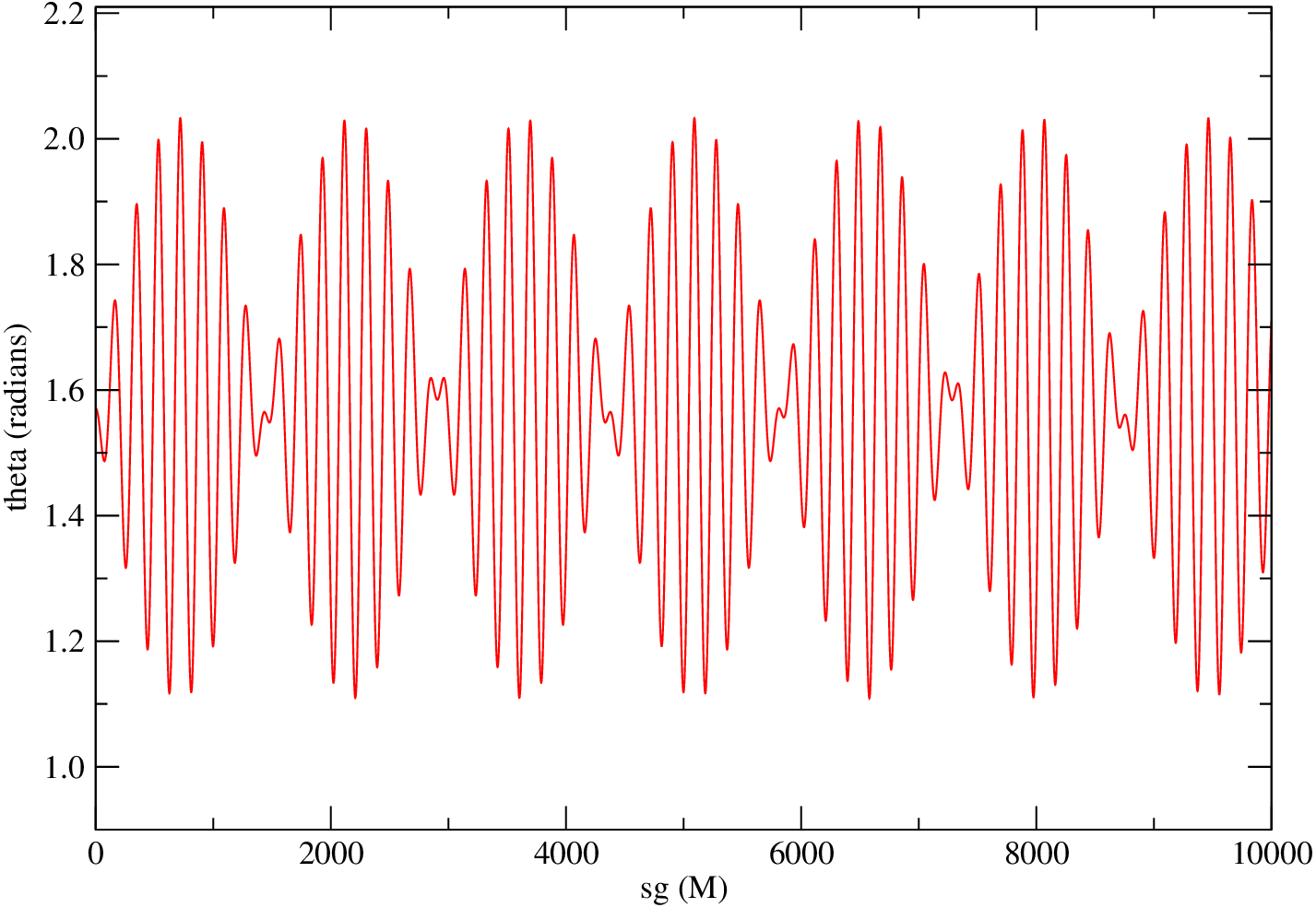}}}
\end{minipage} \\
\caption{\label{fig:theta} The orbital polar angle $\th_s(\sg)$ for various choices of $s/(mr)$,
where $m = 10^{-2} \, M$, $r = 10 \, M$, $\vth = \vph = \pi/4$, and $a = 0.50$.
}
\end{figure*}

\begin{figure*}
\psfrag{sg (M)}[cc][][2.5][0]{\hspace{0.5cm} $\sg$ [M]}
\psfrag{r (M)}[bc][][2.5][0]{\hspace{0.5cm} $r_s(\sg)$ [M]}
\psfrag{dr (M)}[bc][][2.5][0]{\hspace{0.5cm} $\Delta r_s(\tau)$ [M]}
\psfrag{theta (radians)}[bc][][2.5][0]{\hspace{0.5cm} $\th_s(\sg)$ [radians]}
\psfrag{dtheta (radians)}[bc][][2.5][0]{\hspace{0.5cm} $\Delta \th_s(\tau)$ [radians]}
\psfrag{phi (radians)}[bc][][2.5][0]{\hspace{0.5cm} $\ph_s(\sg)$ [radians]}
\psfrag{dphi (radians)}[bc][][2.5][0]{\hspace{0.5cm} $\Delta \ph_s(\tau)$ [radians]}
\begin{minipage}[t]{0.3 \textwidth}
\centering
\subfigure[\hspace{0.2cm} $s/(mr) = 10^{-1}$ (MPD)]{
\label{fig:phi-MPD-s/(mr)=1e-1}
\rotatebox{0}{\includegraphics[width = 6.6cm, height = 5.0cm, scale = 1]{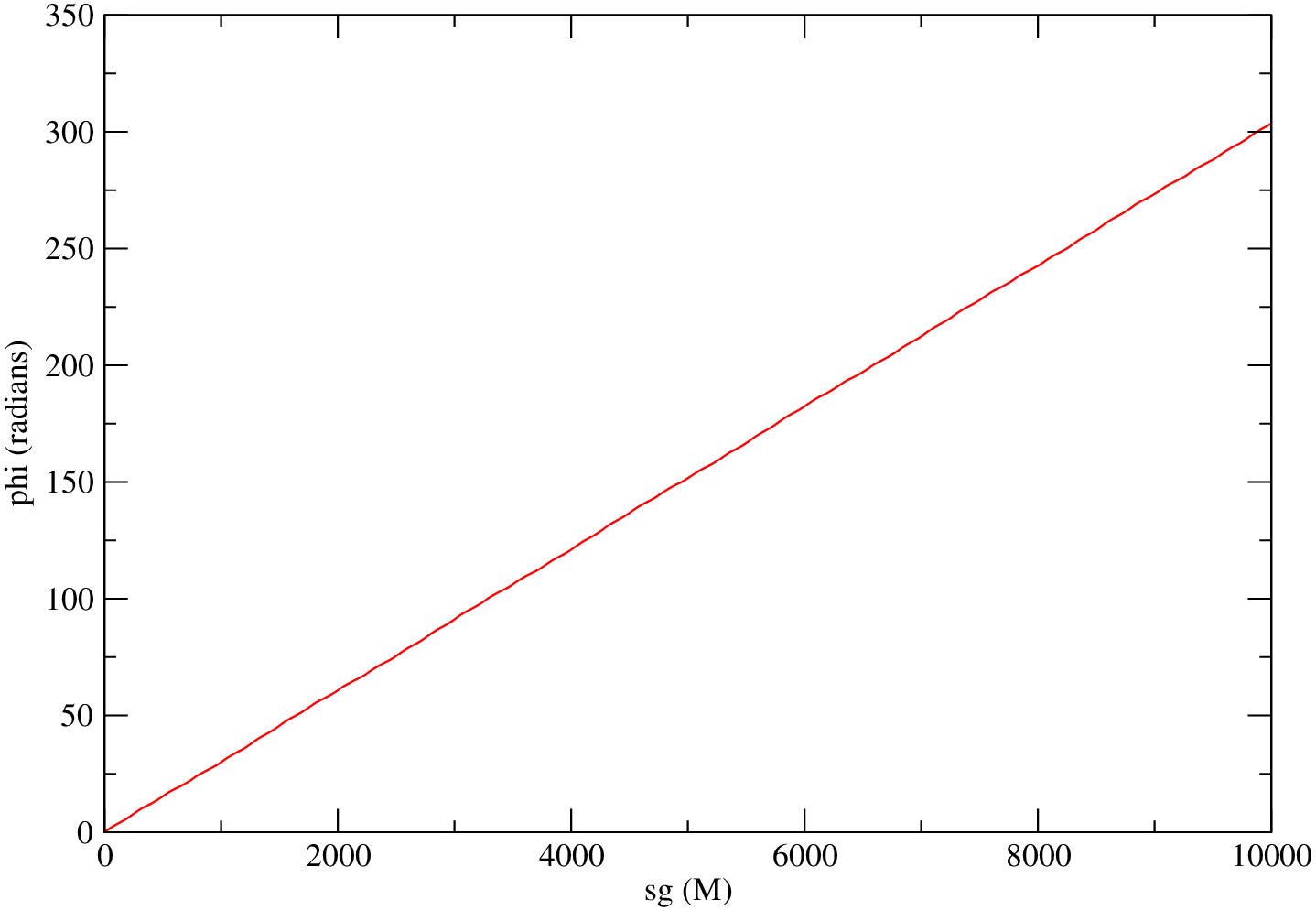}}}
\end{minipage}
\hspace{2.0cm}
\begin{minipage}[t]{0.3 \textwidth}
\centering
\subfigure[\hspace{0.2cm} $s/(mr) = 10^{-1}$ (Linear)]{
\label{fig:phi-lin-s/(mr)=1e-1}
\rotatebox{0}{\includegraphics[width = 6.6cm, height = 5.0cm, scale = 1]{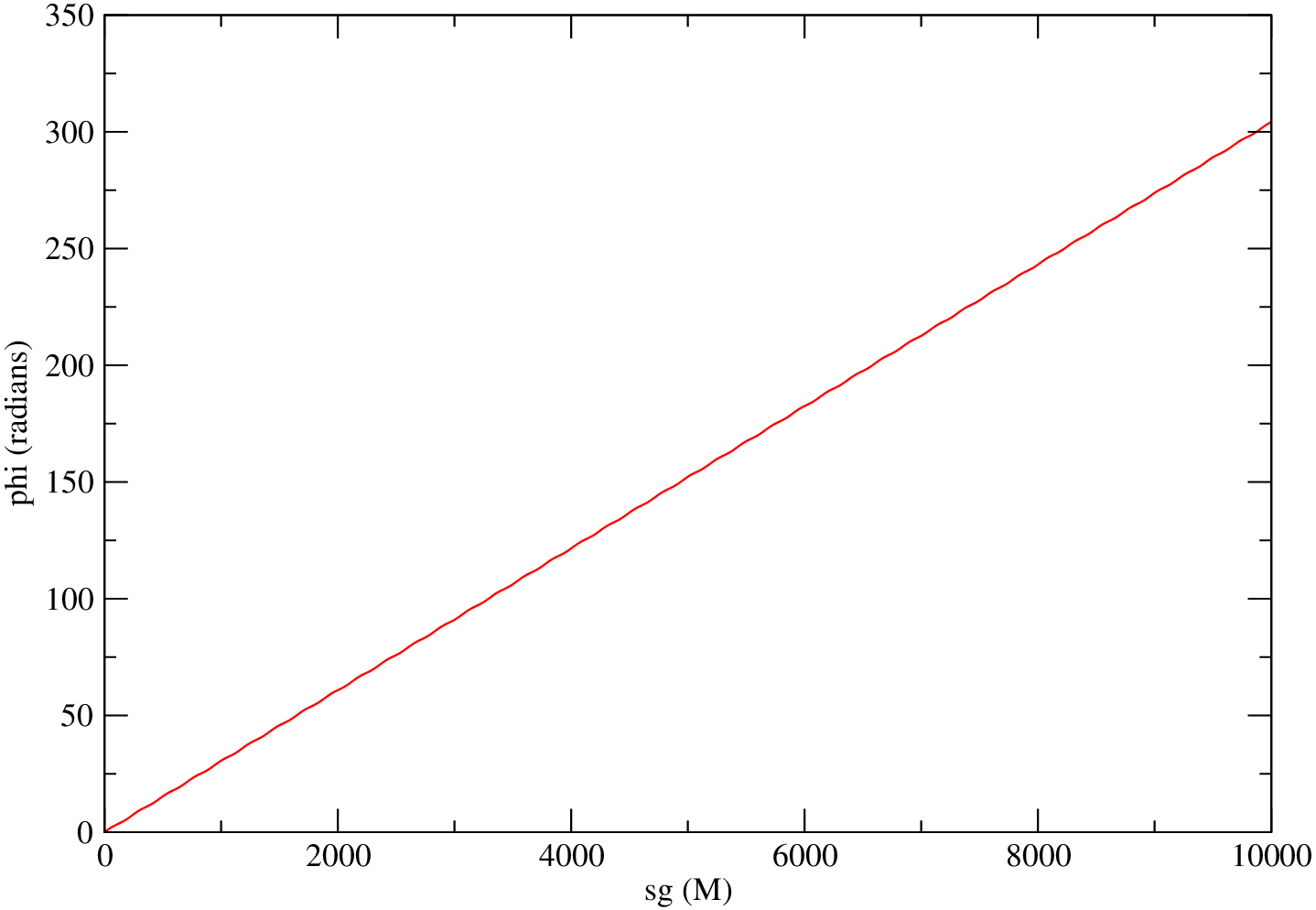}}}
\end{minipage}
\caption{\label{fig:phi} The orbital azimuthal angle $\ph_s(\sg)$,
where $m = 10^{-2} \, M$, $r = 10 \, M$, $\vth = \vph = \pi/4$, and $a = 0.50$.
}
\end{figure*}

We wish to integrate these equations numerically in Kerr spacetime with the same boundary conditions as in the
linear approximation.
Thus, we assume that at $\sg = 0$ the particle starts from $x^\mu(\sg = 0) = \lt(0, r, \pi/2, 0\rt)$
in Boyer-Lindquist co-ordinates.
Moreover, we choose initial conditions for the linear momentum such that $P^\mu(\sg = 0) = m \, u_s^\mu (\sg = 0)$, where
\be
u_s^\mu (\sg = 0) = \lm^\mu{}_{\ze}
\label{MPD-P-ic}
\ee
is given by Eq.~(\ref{tetrad-0}) and corresponds to a stable circular equatorial orbit around the Kerr source.
The initial conditions for the components of the spin vector are given by
\be
S^\mu(\sg = 0) & = & \lm^\mu{}_{\hat{\al}} (\sg = 0) \, S^{\hat{\al}}.
\label{MPD-S-ic}
\ee
Using Eqs.~(\ref{tetrad-1})--(\ref{tetrad-3}) and (\ref{S-vec-comp}) with $S_{\ze} = 0$, we find
\be
S^\mu(\sg = 0) = \lt({m \, L \, \tilde{s} \over r \, A} \, \sin \vph, \, m \, A \, \tilde{s} \, \cos \vph, \,
- {m \, s^* \over r}, \, {m \, E \, \tilde{s} \over r \, A} \, \sin \vph \rt)
\ee
in Boyer-Lindquist co-ordinates.
In this way, Eqs.~(\ref{t-s})--(\ref{ph-s}) should provide the linear approximation to the full solution
of Eqs.~(\ref{dp/dt})--(\ref{dx/dt}), since, as shown in Appendix~\ref{appendix:Fermi}, $\tau = \sg + O(s^2)$.
The results of the integration of the reduced MPD equations for the specific case of a {\em prograde} orbit with
$m = 10^{-2} M$, $\vth = \vph = \pi/4$, $a = 0.5 M$, and $r = 10 M$ are
presented in Figures~\ref{fig:r}--\ref{fig:phi}.
Specifically, the spherical polar co-ordinates $(r, \th, \ph)$ characterizing the orbit of the spinning particle
in the standard Boyer-Lindquist co-ordinates are plotted versus proper time $\sg$ in Figs.~\ref{fig:r}--\ref{fig:phi}, respectively.

The results of the numerical integration of the reduced MPD equations confirm that Eqs.~(\ref{t-s})--(\ref{ph-s})
provide the proper linear approximation to these equations.
It is important to remark that these linear terms primarily exhibit the results of spin-orbit coupling
(cf. Appendix~\ref{appendix:clock-effect}); that is, the rotation of the central source,
though possibly important, is not crucial for the existence of the main phenomena associated with the motion of the
particle in the first-order approximation.
However, the rotation of the source could play a significant role in the difference between prograde and
retrograde motion as in the clock effect discussed in Sec.~\ref{Sec5}.

Beyond the linear approximation, it appears from Figs.~\ref{fig:r}--\ref{fig:phi} that for
radial motion nonlinear spin terms introduce a complex beat phenomenon involving at least two periods.
Of these, the higher frequency is about the same as the beat frequency in the case of polar motion.
Furthermore, nonlinear spin effects are responsible for the reduction of the beat amplitude in the
polar motion.
On the other hand, the azimuthal motion appears to be essentially unaffected by the nonlinearities;
therefore, only the graphs for the highest magnitude of spin considered here are presented in Figure~\ref{fig:phi}.

%
%

\section{Gravitomagnetic Clock Effect}
\label{Sec5}

It follows from the results of the previous section that the analytic formulas involving $s/(mr)$ to first order are a good
approximation to the exact solution of the reduced MPD equations; to go further, higher-order terms in spin must be
taken into account.
On the other hand, the analytic formulas can be safely applied within their theoretical limits of validity to illustrate
a general feature of motion about a rotating source, namely, the gravitomagnetic clock effect \cite{Cohen,Mashhoon2,Bini2,Bini3}.
In its simplest form, the effect involves a spinless ``clock'' on a circular orbit in the equatorial plane of a rotating source.
The difference in the proper periods of two clocks on the same orbit but moving in opposite directions is given by
$\tau_+ - \tau_- \approx 4\pi \, J/(Mc^2)$ for $r \gg 2GM/c^2$.
Here $\tau_+ \lt(\tau_-\rt)$ is the proper period for the prograde (retrograde) motion.
To lowest order, this remarkable effect is independent of Newton's constant of gravitation $G$ and the radius of the orbit.
Moreover, the prograde motion is {\it slower} than the retrograde motion.
For satellites in circular orbit about the Earth, $\tau_+ - \tau_- \approx 10^{-7}$ s; however, measuring this effect will not be simple
\cite{Mashhoon2}.

The purpose of this section is to show that for a spinning ``clock'', the analogous result to first order in $s/(mr)$ is
\be
\tau_+ - \tau_- & \approx & 4\pi \lt[{J \over Mc^2} + 6 \, {s \over mc^2} \lt(\hat{\svec} \cdot \hat{\Jvec}\rt) \rt],
\label{tau+-minus-tau-}
\ee
based on azimuthal closure involving Eq.~(\ref{ph-s}).
This constitutes a generalization of previous results \cite{Faruque1,Bini1,Faruque2} that involved only $\hat{\svec} \cdot \hat{\Jvec} = \pm 1$.

Imagine a spinning particle revolving around a Kerr source in accordance with Eqs.~(\ref{t-s})--(\ref{ph-s}).
After an integral number $p$ of complete orbits with $p \gg 1$, the change in $\ph_{\rm s}$ over a period of time
$p\tau_+ \lt(p \tau_-\rt)$ for the co-rotating (counter-rotating) clock is $2\pi \, p \lt(-2\pi \, p\rt)$.
For $p \gg 1$, we can drop the $\sin \lt(\rho \, \Th\rt)$ term in Eq.~(\ref{ph-s}) in comparison with $\rho \, \Th$; then, we find that the resulting
expressions for $\tau_+$ and $\tau_-$ are such that the one for $\tau_+$ changes over to the one for  $\tau_-$
if $a \rightarrow -a$ and $s \rightarrow -s$.
To first order in $a/r$ and $s/(mr)$, we have
\be
\tau_{\pm} & = & C \, T_{\rm K} \pm {2\pi \over C} \lt(a + 6s^* \, {B^2 \, C^2 \over D^2}\rt),
\label{tau+/-}
\ee
where
\be
B & = & \sqrt{1 - {2M \over r}}, \qquad C \ = \ \sqrt{1 - {3M \over r}}, \qquad D \ = \ \sqrt{1 - {6M \over r}},
\label{C,D}
\ee
$T_{\rm K} = 2\pi/|\OmK|$ is the Keplerian period and $s^*$ given by Eq.~(\ref{s-scales})
is proportional to the component of the specific spin of the ``clock'' along the axis of rotation of the source.
Let us note that $\tau_+ - \tau_-$ computed from Eq.~(\ref{tau+/-}) for $r \gg 2M$ would result in Eq.~(\ref{tau+-minus-tau-}).

It is interesting to remark here that in the equatorial plane of Kerr spacetime,
\be
t_+ - t_- & = & 4\pi \, a,
\label{t+-minus-t-}
\ee
which is another manifestation of the special gravitomagnetic temporal structure around the rotating source.
Here $t_+$ and $t_-$ are the periods of geodesic motion around the same circular orbit according to the static
inertial observers at spatial infinity.
Substituting Eq.~(\ref{tau+/-}) in Eq.~(\ref{t-s}), we find along the same lines as before that the corresponding
result for the motion of a spinning particle to first order in $s/(mr)$ and $a/r$ is $t_\pm = T_{\rm K} \pm 2\pi\lt(a + 6s^* C^2/D^2\rt)$,
so that
\be
t_+ - t_- & = & 4\pi \lt(a + 6s^* \, {C^2 \over D^2}\rt).
\label{t+-minus-t-1}
\ee

These results are in qualitative agreement with previous work in this direction \cite{Faruque1,Bini1,Faruque2} that has
been restricted to {\em circular} orbits for the motion of the spinning particle around the Kerr source with the
corresponding limitation that the spin of the particle has to be parallel or antiparallel to the Kerr axis \cite{Abramowicz}.
In this connection, it is important to remark here that Eqs.~(\ref{t-s})--(\ref{ph-s}) explicitly forbid a circular
equatorial orbit with $\cos \vth = \pm 1$.
Thus, a direct quantitative comparison with previous work is not possible in this case.
The contribution of spin to the clock effect is further elucidated in Appendix~\ref{appendix:clock-effect}.
%

\section{Conclusion}
\label{Sec6}

We have developed in this paper a general first-order approximation scheme for the effect of spin on the motion of an extended spinning test particle in a
gravitational field in accordance with the MPD equations that are based on the multipole expansion method.
We have neglected the influence of quadrupole and higher moments on the motion.
For the astrophysically interesting case of motion in the field of a central source, we must have $s/(mcr) \ll 1$,
where $s/(mc)$ is the M{\o}ller radius of the extended test particle.
Thus the influence of the spin of the particle on its motion is treated to linear order throughout.
It turns out that in this approximation the spin is parallel transported along the nongeodesic path of the test particle.
To illustrate our general approach, the motion of spinning test particles along nearly circular equatorial orbits in the exterior
gravitational field of a Kerr source has been investigated in detail and the results of the first-order
approximation scheme have been numerically compared with the corresponding solution of the full MPD equations.
As expected, the results are essentially identical for $s/(mcr)$ sufficiently small compared to unity;
moreover, we have described the deviations that appear when $s/(mcr)$ is not so small.

The astrophysical implications of our results have been briefly mentioned in Sec.~\ref{Sec3}.
For the motion of the Earth about the Sun, for instance, the spin-induced deviations from Keplerian motion
are too small to be observationally significant at present; however, the spin-dependent terms could become
important in binary pulsars \cite{Iorio,Barker}.
Furthermore, we have elucidated the gravitomagnetic clock effect for spinning particles in accordance
with our first-order approximation method.

\section{Acknowledgements}
\label{Sec7}

DS wishes to thank Nader Mobed of the University of Regina for financial support via an NSERC Discovery Grant.

\begin{appendix}

\section{MPD Equations in Fermi Normal Coordinates}
\label{appendix:Fermi}
\renewcommand{\theequation}{A.\arabic{equation}}
\setcounter{subsection}{0}
\setcounter{equation}{0}

Consider a Fermi normal coordinate system that is constructed about the reference geodesic $x^\mu(\tau)$ and is based upon the
parallel-propagated orthonormal frame $\lm^\mu{}_{\Cal}$.
Specifically, along the worldline of the spinning particle $x_s^\mu$, let us imagine an event $Q_s$ with Fermi coordinates
$X^\mu = \lt(T,\Xvec\rt)$; then, there exists a unique spacelike geodesic of proper length $l$ that connects $Q_s$ orthogonally
to the reference geodesic $x^\mu$ at the event $Q_0$ with proper time $\tau$ such that
\be
T & = & \tau, \qquad X_i \ = \ l \, n_\mu \, \lm^\mu{}_{\Ci},
\label{A1}
\ee
where $n_\mu = \lt(dx^\mu/dl\rt)_{Q_0}$ is the unit vector tangent to the unique spacelike geodesic at $Q_0$ such that $n_\mu \, u^\mu = 0$.
Thus the spinless reference particle following $x^\mu(\tau)$ permanently occupies the spatial origin of this Fermi coordinate system
and has Fermi coordinates $\lt(T,\vec{0}\rt)$, where $T = \tau$.
The spacetime metric in Fermi coordinates is given by
\be
g_{00} & = & -1 - R_{\ze \Ci \ze \Cj}(T) \, X^i \, X^j + \cdots,
\label{A2}
\nl
g_{0i} & = & -{2 \over 3} \, R_{\ze \Cj \Ci \Ck}(T) \, X^j \, X^k + \cdots,
\label{A3}
\nl
g_{ij} & = & \dl_{ij} - {1 \over 3} \, R_{\Ci \Ck \Cj \Cl}(T) \, X^k \, X^l + \cdots,
\label{A4}
\ee
where $R_{\Cal \Cbt \Cgm \Cdl}$, given by the projection of the Riemann curvature tensor on the tetrad
frame of the spinless reference particle, is in fact the Riemann tensor in the Fermi coordinate system evaluated at its spatial origin.
The Fermi coordinate system is admissible within a cylindrical spacetime region around the worldline of the reference geodesic such that
$|\Xvec| < {\cal R}_a(T)$, where ${\cal R}_a$ is a certain minimum radius of curvature of spacetime.

The Mathisson-Papapetrou-Dixon equations of motion of the spinning particle within the framework of our first-order approximation scheme
(based on $s \ll mr$) can be expressed in the Fermi coordinate system as
\be
{DU^\mu \over d\sg} & = & {\cal A}^\mu, \qquad {\cal A}^\mu \ = \ -{1 \over 2m} \, R^\mu{}_{\gm \al \bt} \, U^\gm \, S^{\al \bt},
\label{A5}
\nl
{DS^{\mu \nu} \over d\sg} & = & 0, \qquad S_{\mu \nu} \, U^\nu \ = \ 0,
\label{A6}
\ee
where $U^\mu = dX^\mu/d\sg$ is the four-velocity of the spinning particle.
Let us write this as
\be
U^\mu & = & \Gm\lt(1,\Vvec\rt), \qquad \Gm \ = \ {dT \over d\sg}, \qquad \Vvec \ = \ {d\Xvec \over dT}.
\label{A7}
\ee
Then,
\be
\Gm^{-2} & = & -g_{00} - 2 \, g_{0i} \, V^i - g_{ij} \, V^i \, V^j \ > \ 0,
\label{A8}
\ee
since $U^\mu$ is a timelike unit vector.
To characterize the order of various quantities in accordance with our perturbation method, it is convenient to write
\be
\Xvec(T) & = & O(s), \qquad \Vvec(T) \ = \ O(s), \qquad {1 \over \Gm^2} \ = \ 1 + O(s^2),
\label{A9}
\ee
and so on; in particular, it follows that $\sg = T + O(s^2)$.
We have already discussed the consequences of Eq.~(\ref{A6}) in Sec.~\ref{Sec2}; therefore, we concentrate here on Eq.~(\ref{A5}).
We find \cite{Chicone1,Chicone4}
\be
{d^2 X^i \over dT^2} + \lt(\Gm^i_{\al \bt} - \Gm^0_{\al \bt} \, V^i\rt) {dX^\al \over dT} \, {dX^\bt \over dT} \ = \ {1 \over \Gm^2} \lt({\cal A}^i - {\cal A}^0 \, V^i\rt),
\label{A10}
\ee
where
\be
{\cal A}^0 & = & 0, \qquad {\cal A}^i \ = \ {1 \over 2m} \, R_{\ze \Ci \Cj \Ck}(T) \, S^{\Cj \Ck}.
\label{A11}
\ee
We note that the Mathisson-Papapetrou acceleration is given by
\be
{\cal A}^i & = & {1 \over m} \, {\cal H}_{ij} \, S^{\Cj}
\label{A12}
\ee
based on the definitions introduced in Sec.~\ref{Sec2}.
Moreover, of the connection coefficients in Eq.~(\ref{A10}) only $\Gm^i_{00} = {\cal E}_{ij} \, X^j$ makes a
non-negligible contribution.
Thus taking due account of our first-order approximation scheme, we find that Eqs.~(\ref{A10})--(\ref{A12}) reduce to
\be
{d^2X^i \over dT^2} + {\cal E}_{ij} \, X^j & = & {\cal A}^i
\label{A13}
\ee
with the boundary conditions that $\Xvec = \vec{0}$ and $\Vvec = \vec{0}$ at $T = 0$.
We now turn to the general solution of Eq.~(\ref{A13}), which may be expressed as
\be
{d \Psi \over dT} & = & \lt(
\begin{array}{cc}
0 & I \\
-{\cal E} & 0
\end{array} \rt) \Psi + \chi,
\label{A14}
\ee
where $I$ is the $3 \times 3$ unit matrix and
\be
\Psi & = & \lt(
\begin{array}{c}
\Xvec \\ \Vvec
\end{array} \rt), \qquad
\chi \ = \ \lt(
\begin{array}{c}
\vec{0} \\ {\bf {\cal A}}
\end{array} \rt).
\label{A15}
\ee

Let $\psi$ be a general solution of the homogeneous (Jacobi) system with $\chi = \vec{0}$; then,
\be
\psi & = & \sum_{i = 1}^6 c_i \, \psi_i,
\label{A16}
\ee
where the $c_i$, $i = 1, 2, \cdots, 6$, are arbitrary constants that physically correspond to the initial position
and velocity of a free particle following the Jacobi equation and $\psi_i$, $i = 1, 2, \cdots, 6$, form a fundamental
set of solutions of the Jacobi system.
According to the method of variation of parameters \cite{Chicone2}, we seek a solution of the inhomogeneous Eq.~(\ref{A14})
by assuming that $c_i \rightarrow C_i(T)$, i.e., we let
\be
\Psi & = & \sum_{i = 1}^6 C_i(T) \, \psi_i.
\label{A17}
\ee
Substitution of Eq.~(\ref{A17}) in Eq.~(\ref{A14}) results in
\be
\sum_{i = 1}^6 {dC_i \over dT} \, \psi_i & = & \chi.
\label{A18}
\ee
Consider a $6 \times 6$ matrix $\Phi$ that is the fundamental matrix solution of the homogeneous Jacobi system and contains
$\psi_1, \psi_2, \cdots, \psi_6$ as its column vectors.
Then $\Psi = \Phi \, {\cal C}$, where ${\cal C}$ is a column vector with $C_1, C_2, \cdots, C_6$ as its elements,
and Eq.~(\ref{A18}) can be written as
\be
{d{\cal C} \over dT} & = & \Phi^{-1} \, \chi.
\label{A19}
\ee
With the initial conditions that $\Psi = {\cal C} = 0$ at $T = 0$, the solution of Eq.~(\ref{A14}) is
\be
\Psi(T) & = & \Phi(T) \int_0^T \, \Phi^{-1}(T') \, \chi(T') \, dT'.
\label{A20}
\ee

In principle, the trajectory of the spinning particle can thus be determined in the Fermi coordinate system.
To express this trajectory in terms of the original background coordinate system, it is in general necessary to have
the explicit coordinate transformation between the two systems of coordinates.
This turns out to be possible only in very special situations \cite{Bini4,Chicone5}.
However, our first-order approximation scheme makes it possible to proceed as follows.
The deviation vector $\dl x^\mu$ may be written as $\dl x^\mu \approx l \, n^\mu$, since $l \ll {\cal R}_a$.
Therefore, regarding $\dl x^\mu$ as a vector field along the reference geodesic, we can write $\dl x^\mu = X^i(\tau) \, \lm^\mu{}_{\Ci}$,
where the Fermi temporal coordinate in the solution of Eq.~(\ref{A13}), namely, $\Xvec(T)$, has been replaced by $\tau$.
It follows that $x_s^\mu = x^\mu + X^i \, \lm^\mu{}_{\Ci}$; in this way, we recover Eqs.~(\ref{delta-x-mu}) and (\ref{x-s=}) of Sec.~\ref{Sec2}.

\section{Heuristic Interpretation of the Clock Effect for Spinning Particles}
\label{appendix:clock-effect}
\renewcommand{\theequation}{B.\arabic{equation}}
\setcounter{subsection}{0}
\setcounter{equation}{0}

The purpose of this appendix is to provide an intuitive physical understanding for the appearance of terms proportional
to $s^*$ in the discussion of the gravitomagnetic clock effect in Sec.~\ref{Sec5}.
The existence of such a term was first pointed out in Ref.~\cite{Faruque1} and has been further studied in
\cite{Bini1} and \cite{Faruque2}.
The approach adopted here is based on the gravitoelectromagnetic (``GEM'') analogy; we follow here the basic
conventions of Ref.~\cite{Mashhoon3}.

Consider a spinning particle in orbit about a central {\em nonrotating} source.
In the rest frame of the particle, the central source revolves around the particle and this motion generates a gravitomagnetic field.
The resulting spin-gravitomagnetic field interaction, just as in the electromagnetically analogous
case of the hydrogen atom, must be corrected by taking due account of Thomas precession, which in the
gravitational case becomes the de Sitter-Fokker (``geodetic'') precession.
The combined interaction due to these terms is of the spin-orbit coupling form \cite{Barker}.
Specifically, in the rest frame of the spinning particle, the gravitomagnetic field $\vec{B}_g$ is
proportional to $-\lt(\vec{v}/c\rt) \times \vec{E}_g$, where $\vec{E}_g = GM \, \vec{r}/r^3$.
The corresponding contribution to the Hamiltonian, i.e. the analogue of $-\vec{\mu} \cdot \vec{B}$, would be \cite{Mashhoon3}
\be
{1 \over c} \, \vec{s} \cdot \vec{B}_g & = & \xi_0 \lt({GM \over mc^2 \, r^3}\rt) \vec{s} \cdot \vec{L},
\label{s.B}
\ee
where $\vec{L} = m \rvec \times \vec{v}$ and $\xi_0$ is a numerical factor that is expected to be of order unity.
Furthermore, the geodetic precession frequency is given by
\be
\vec{\omega}_g & = & {3 \over 2} \lt({GM \over mc^2 \, r^3}\rt) \vec{L},
\label{om-g}
\ee
hence the corresponding contribution to the Hamiltonian would be
\be
\vec{s} \cdot \vec{\omega}_g & = & {3 \over 2} \lt({GM \over mc^2 \, r^3}\rt) \vec{s} \cdot \vec{L}.
\label{s.omega}
\ee
Therefore, the net Hamiltonian for the motion of the particle may be taken to be of the form
\be
H & = & {p^2 \over 2m} - {GM \, m \over r} + \xi \lt({GM \over mc^2 \, r^3}\rt) \vec{s} \cdot \vec{L}
\label{H}
\ee
in this simple GEM model.
Here,
\be
\xi & = & \xi_0 + {3 \over 2}.
\label{xi}
\ee
In Eq.~(\ref{H}), the spin-orbit term is a small relativistic correction to the Newtonian dynamics and turns out to be
the source of the spin-dependence of the clock effect \cite{Faruque2}.

It follows from Hamilton's equations of motion for Eq.~(\ref{H}) that
\be
{d \rvec \over dt} & = & {\pvec \over m} + \xi \lt({GM \over mc^2 \, r^3}\rt) \vec{s} \times \rvec,
\label{dr-dt}
\nl
{d \pvec \over dt} & = & -\lt({GM \, m \over r^3}\rt) \rvec + 3 \, \xi \lt({GM \over mc^2 \, r^5}\rt) \lt(\vec{s} \cdot \vec{L}\rt) \rvec
- \xi \lt({GM \over mc^2 \, r^3}\rt) \pvec \times \vec{s}.
\label{dp-dt}
\ee
Thus the canonical momentum can be expressed as
\be
\pvec & = & m \, \vec{v} + \xi \lt({GM \over c^2 \, r^3}\rt) \rvec \times \vec{s}
\label{p}
\ee
and the equation of motion of the spinning particle is then given by the substitution of Eq.~(\ref{p}) in Eq.~(\ref{dp-dt}).
To first order in the spin-dependent contribution, the result is
\be
{d^2 \rvec \over dt^2} + \lt({GM \over r^3}\rt) \rvec & = & {\vec{F} \over m},
\label{eq-motion}
\ee
where $\vec{F}$ is the spin-dependent force given by
\be
\vec{F} & = & \xi \lt({GM \over c^2 \, r^3}\rt) \lt\{{3 \over r} \, {dr \over dt} \lt(\rvec \times \vec{s}\rt)
- 2 \lt(\vec{v} \times \vec{s}\rt) + {3 \over r^2} \lt[\lt(\rvec \times \vec{v}\rt) \cdot \vec{s}\rt]\rvec \rt\}.
\label{F}
\ee

Consider an initially circular Keplerian orbit of radius $r$ in the equatorial $(x,y)$ plane.
At $t = 0$, the orbit is perturbed by the spin-dependent force
\be
\vec{F} & = & \xi \, \OmK \lt({GM \over c^2 \, r^3}\rt) \lt[\lt(s \, \cos \vth\rt)\rvec + 2 \lt(\vec{s} \cdot \rvec\rt)\hat{\vec{z}} \rt],
\label{F1}
\ee
where $\vec{s} = \lt(s, \vth, \vph\rt)$ in spherical polar co-ordinates and
\be
\rvec & = & r \lt[\lt(\cos \OmK t\rt) \hat{\vec{x}} + \lt(\sin \OmK t\rt) \hat{\vec{y}} \rt].
\label{r}
\ee
Let us write the equation of the perturbed orbit in cylindrical co-ordinates as
\be
r_s & = & r \lt(1 + f\rt), \quad \phi_s \ = \ \OmK t + q, \quad z_s \ = \ r \, h,
\label{perturbations}
\ee
where $f$, $q$, and $h$ denote spin-dependent perturbations.
Imposing the boundary conditions that at $t=0$,
\be
f & = & q \ = \ h \ = \ 0, \quad {df \over dt} \ = \ {dq \over dt} \ = \ {dh \over dt} \ = \ 0,
\label{perturb-bc's}
\ee
we find, using the method developed in Ref.~\cite{Mashhoon4}, that
\be
r_s & = & r + \xi \sqrt{GM \over c^2 \, r} \, s^* \lt(1 - \cos \OmK t \rt),
\label{r-1}
\nl
\ph_s & = & \OmK t - {2 \over c} \, \xi \, \OmK \, s^* \lt(\OmK t - \sin \OmK t\rt),
\label{ph-1}
\nl
z_s & = & \xi \sqrt{GM \over c^2 \, r} \, \tilde{s} \lt[\sin \lt(\OmK t - \vph\rt) \OmK t + \sin \vph \, \sin \OmK t\rt].
\label{z-1}
\ee

It is interesting to compare Eqs.~(\ref{r-1})--(\ref{z-1}) with Eqs.~(\ref{t-s})--(\ref{ph-s});
in fact, to lowest relativistic order, we find from the latter equations the same results as Eqs.~(\ref{r-1}) and (\ref{ph-1})
with $\xi = 3$.
Moreover, with $\zeta \Th \approx \OmK t, \, \zeta \approx 1 + \lt(3/2\rt)\bt^2$, and $\bt^2 \approx GM/(c^2 \, r)$,
Eq.~(\ref{th-s}) implies that
\be
\th_s & \approx & {\pi \over 2} - {3 \over 2} \sqrt{GM \over c^2 \, r} \, {\tilde{s} \over r}
\lt[\sin \lt(\OmK t - \vph\rt) \OmK t + \sin \vph \, \sin \OmK t\rt].
\label{th-1}
\ee
We note that $z_s = r_s \, \cos \vth_s$, so that one obtains from Eqs.~(\ref{t-s})--(\ref{ph-s}) that
\be
z_s & \approx & {3 \over 2} \sqrt{GM \over c^2 \, r} \, \tilde{s} \lt[\sin \lt(\OmK t - \vph\rt) \OmK t + \sin \vph \, \sin \OmK t\rt],
\label{z-2}
\ee
which is smaller by a numerical factor of $1/2$ than the expression given by Eq.~(\ref{z-1}) with $\xi = 3$.
Nevertheless, it is rather remarkable that our simple model, based on the GEM analogy, predicts the main qualitative features of
the perturbed orbit.
In particular, it follows from Eq.~(\ref{ph-1}) that
\be
t_\pm & = & T_K \pm 4 \pi \, \xi \, s^*;
\label{t-pm}
\ee
hence, for $\xi = 3$ the spin part of the clock effect is recovered.

We have thus far ignored the proper rotation of the central source.
The rotation of the source would generate a gravitomagnetic field and the spin of the test mass naturally couples
to this gravitomagnetic field as described in detail in Ref.~\cite{Chicone1}.
It turns out, however, that the corresponding {\em dominant} ``hyperfine'' coupling term is independent of the sense of the orbit
and hence does not contribute to the main spin-dependent gravitomagnetic clock effect under consideration here.

Finally, a remark is in order here regarding the fact that with $\xi = 3$ and $\xi = \xi_0 + 3/2$, we have $\xi_0 = 3/2$,
so that the $\vec{s} \cdot \vec{B}_g/c$ and geodetic terms, given respectively by Eqs.~(\ref{s.B}) and (\ref{s.omega}), contribute
equally to the net spin-orbit coupling term in the classical Hamiltonian (\ref{H}).
As noted in Ref.~\cite{Faruque2}, a similar result has been obtained in the treatment of a Dirac particle in the
Schwarzschild field \cite{Lee,Hehl,Audretsch,Varju}.

\end{appendix}

\end{document}